\documentclass{article}

\usepackage{arxiv}

\usepackage[utf8]{inputenc} 
\usepackage[T1]{fontenc}    
\usepackage{hyperref}       
\usepackage{url}            
\usepackage{booktabs}       
\usepackage{amsfonts}       
\usepackage{nicefrac}       
\usepackage{microtype}      
\usepackage{lipsum}

\usepackage{color,setspace}
\usepackage{caption}
\usepackage{amsfonts,amsmath,amssymb,amsthm,mathtools}
\usepackage{epsfig}
\usepackage{bm}
\usepackage[square, numbers, sort&compress]{natbib}
\usepackage{multirow}
\usepackage{flushend}
\usepackage{epstopdf}
\usepackage{graphicx}
\usepackage{booktabs}
\usepackage{enumerate}

\DeclareMathOperator*{\argmin}{arg\,min}

\newtheoremstyle{mynewtheorem}
{3pt}
{3pt}
{\itshape}
{}
{\bfseries}
{.}
{.5em}
{\thmname{#1} \thmnumber{#2}}%
\theoremstyle{mynewtheorem}

\newtheorem{myprop}{Proposition}
\newtheorem{mylemma}{Lemma}

\newcommand{\dotp}{\dot{\mathbf p}}
\newcommand{\ddotp}{\ddot{\mathbf p}}
\newcommand{\dotv}{\dot{\mathbf v}}
\newcommand{\ddotv}{\ddot{\mathbf v}}

\newcommand{\eqdef}{:=}

\def\tsc#1{\csdef{#1}{\textsc{\lowercase{#1}}\xspace}}
\tsc{WGM}
\tsc{QE}
\tsc{EP}
\tsc{PMS}
\tsc{BEC}
\tsc{DE}

\title{Minimum Mean-Squared-Error Autocorrelation Processing in Coprime Arrays}
\author{Dimitris G. Chachlakis\\
  Dept. of Electrical and Microelectronic Engineering\\
  Rochester Institute of Technology\\
  Rochester, NY 14623\\
  \texttt{dimitris@mail.rit.edu}
   \And
  Tongdi Zhou\\
  Dept. of Electrical and Computer Engineering\\
  Temple University\\
  Philadelphia, PA 19122\\
  \texttt{tzhou@temple.edu}
   \And
  Fauzia Ahmad\\
  Dept. of Electrical and Computer Engineering\\
  Temple University\\ Philadelphia, PA 19122\\
  \texttt{fauzia.ahmad@temple.edu}
  \And 
  Panos P. Markopoulos\thanks{Corresponding author.\newline Preliminary results of this work were presented at the IEEE International Workshop on Computational Advances in Multi-sensor Adaptive Processing Workshop, Cura\c{c}ao, Dutch Antilles, December 2017 \cite{DGC1} and IEEE International Conference on Acoustics, Speech and Signal Processing, Alberta, Canada, April 2018 \cite{DGC2}.}\\
  Dept. of Electrical and Microelectronic Engineering\\
  Rochester Institute of Technology\\
  Rochester, NY 14623 \\
  \texttt{panos@rit.edu} 
}

\begin{document}
\maketitle

\begin{abstract}
Coprime arrays enable Direction-of-Arrival (DoA) estimation of an increased number of sources. To that end, the receiver estimates the autocorrelation matrix of a larger virtual uniform linear array (coarray), by applying selection or averaging to the physical array's autocorrelation estimates, followed by spatial-smoothing. Both selection and averaging have been designed under no optimality criterion and attain arbitrary (suboptimal) Mean-Squared-Error (MSE) estimation performance. In this work, we design a novel coprime array receiver that estimates the coarray autocorrelations with Minimum-MSE (MMSE), for any probability distribution of the source DoAs. Our extensive numerical evaluation illustrates that the proposed MMSE approach returns superior autocorrelation estimates which, in turn, enable higher DoA estimation performance compared to standard counterparts. 
\end{abstract}


\section{Introduction}
Coprime arrays (CAs) are non-uniform linear arrays with element locations determined by a pair of distinct coprime numbers.  
CAs are a special class of sparse arrays \cite{PP1, PP9} which
are often preferred due to their desirable bearing properties (e.g., enhanced degrees of freedom and closed-form expressions for element-locations).
CAs have attracted significant research interest over the past years and have been successfully employed in applications such as Direction-of-Arrival (DoA) estimation \cite{FAUZHIA1,AMIN0,AMIN2,AMIN3,PP3,PP4,CHUNLIU1,CHUNLIU3,CHUNLIU4, FA8,ZTAN,MWANG,GOODMAN}, beamforming  \cite{PP5,CZHOU2, Buck17}, interference localization and mitigation in satellite systems \cite{AMIN6}, and space-time adaptive processing \cite{CHUNLIU2}, to name a few.  
More recently, scholars have explored/employed CAs for underwater localization \cite{underwater1, underwater2}, channel estimation in MIMO communications via tensor decomposition \cite{tensors}, receivers on moving platforms which further increase degrees of freedom \cite{coprimeMoving,Krolik1}, and receivers capable of two-dimensional DoA estimation.

In standard DoA with CAs \cite{PP1}, the receiver conducts a series of intelligent processing steps and assembles an autocorrelation matrix which corresponds to a larger virtual Uniform Linear Array (ULA), known as the coarray. 
Accordingly, CAs enable the identification of more sources than physical sensors compared to equal-length ULAs.     
Processing at a coprime array receiver commences with the estimation of the nominal (true) physical-array autocorrelations based on a collection of received-signal snapshots.
The receiver processes the estimated autocorrelations so that each coarray element is represented by one autocorrelation estimate. 
Next, the processed autocorrelations undergo spatial smoothing, forming an autocorrelation matrix estimate which corresponds to the coarray. 
Finally, a DoA estimation approach, such as the MUltiple SIgnal Classification (MUSIC) algorithm, can be applied on the resulting autocorrelation matrix estimate for identifying the source directions. 

At the autocorrelation processing step, the estimated autocorrelations are commonly processed by \emph{selection combining} \cite{PP1}, retaining only one autocorrelation sample for each coarray element.
Alternatively, an autocorrelation estimate for each coarray element is obtained by \emph{averaging combining} \cite{CHUNLIU1} all available sample-estimates corresponding to a particular coarray element.
The two methods coincide in Mean-Squared-Error (MSE) estimation performance when applied on the nominal physical-array autocorrelations--which the receiver could only estimate with asymptotically large number of received-signal snapshots. 
In practice, due to a finite number of received-signal snapshots available at the receiver and the fact that these methods have been designed under no optimality criterion, the estimated autocorrelations diverge from the nominal ones and attain arbitrary MSE performance. In this case, the two methods no longer coincide in MSE estimation performance. It was recently shown in \cite{DGC1} that averaging combining attains superior estimation performance compared to selection combining with respect to the MSE metric. 

Motivated by prior works which treat angular variables as statistical random variables \cite{randomvDoA0,randomvDoA1}, in this work, we make the mild assumption that the DoAs are independent and identically distributed random variables and design a novel coprime array receiver equipped with a linear autocorrelation combiner which is designed under the Minimum-MSE (MMSE) optimality criterion. The proposed MMSE combiner minimizes, in the mean (i.e., for any configuration of DoAs), the error in estimating the physical-array autocorrelations with respect to the MSE metric. Moreover, we review the MSE expressions of selection and averaging combining of \cite{DGC1} and, for the first time, offer formal mathematical proofs for these expressions. Finally, we conduct extensive numerical studies and compare the performance of the proposed MMSE combiner to existing counterparts, with respect to autocorrelation estimation error and DoA estimation.

The rest of this paper is organized as follows. In Section \ref{SEC: SIGNAL}, we present the signal model and state the problem of interest. 
In Section \ref{SEC: LITERATURE}, we review existing selection and averaging autocorrelation combining methods for coprime arrays, providing their closed-form MSE expressions  \cite{DGC1}, and offering formal mathematical proofs for these expressions. 
We present the proposed MMSE autocorrelation combining approach in Section \ref{SEC: PROPOSED}.
Next, in Section \ref{SEC:NUMERICAL}, we conduct extensive numerical performance evaluations of the proposed combining approach and compare against existing counterparts.
Conclusions are drawn in Section \ref{SEC:CONCLUSIONS}.

\section{Signal Model}
\label{SEC: SIGNAL}
Consider coprime naturals $(M,N)$ such that $M<N$. A coprime array equipped with $L=2M+N-1$ antenna elements is formed by overlapping a ULA with $N$ antenna elements at positions  $ p_{M,i} = (i-1)Md$, $i=1,2,\ldots,N$, and a ULA equipped with $2M-1$ antenna elements at positions $ p_{N,i} = iNd$, $i=1,2,\ldots, 2M-1$. The reference unit-spacing $d$ is typically set to one-half wavelength at the operating frequency.
The positions of the $L$ elements of the coprime array are described by the element-location vector $\mathbf p \eqdef \mathrm{sort}([p_{M,1},\ldots, p_{M,N}, p_{N,1}, \ldots, p_{N,2M-1}]^\top)$, where  $\mathrm{sort}(\cdot)$
sorts the entries of its vector argument in ascending order and the superscript `$\top$' denotes matrix transpose. We assume that narrowband signals impinge on the array from $K < MN+M$ sources with propagation speed $c$ and carrier frequency $f_c$. 
Assuming far-field conditions, a signal from source $k \in \{ 1,2,\ldots, K \}$ impinges on the array from direction $\theta_k \in (-\frac{\pi}{2}, \frac{\pi}{2}]$
with respect to the broadside. The array response vector for source $k$ is $ \mathbf s(\theta_k) \eqdef   \left[ v(\theta_k)^{[\mathbf p]_1},    \ldots, v(\theta_k)^{[\mathbf p]_L} \right]^\top \in \mathbb C^{L \times 1},$ with $v (\theta) \eqdef \mathrm{exp}\left(\frac{-j 2 \pi f_c}{c} \mathrm{sin}(\theta) \right)$ for every $\theta \in (-\frac{\pi}{2}, \frac{\pi}{2}] $.
Accordingly, the $q$th collected received-signal snapshot is of the form 
\begin{align}
\mathbf y_q = \sum_{k=1}^K \mathbf s(\theta_k) \xi_{q,k} + \mathbf n_q \in \mathbb C^{L \times 1},
\label{snapshots}
\end{align} 
where $\xi_{q,k}\sim \mathcal{CN}(0,d_k)$ is the $q$th symbol for source $k$ (power-scaled and flat-fading-channel processed) and $\mathbf n_q \sim \mathcal {CN} (\mathbf 0_L, \sigma^2 \mathbf I_L)$ models Additive White Gaussian Noise (AWGN). 
We make the common assumptions that  the random variables are statistically independent across different snapshots and symbols from different sources are independent of each other and of every entry of $\mathbf n_{q}$.
The received-signal autocorrelation matrix is given by 
\begin{align}
\mathbf R_{y} \eqdef \mathbb E\{ \mathbf y_q \mathbf y_q^H \} = \mathbf S ~\mathrm{diag}(\mathbf d)~ \mathbf S^H + \sigma^2 \mathbf I_L,
\label{eq:Ry}
\end{align}
where $\mathbf d \eqdef [d_{1}, d_{2}, \ldots, d_{K}]^\top \in \mathbb R_+^{K \times 1}$ is the source-power vector and $\mathbf S \eqdef [\mathbf s(\theta_1), $ $ \mathbf s(\theta_2), $ $\ldots, \mathbf s(\theta_K)] \in \mathbb C^{L \times K}$ is the array-response matrix.
We define
\begin{align}
\mathbf r \eqdef \mathrm{vec} ( \mathbf R_{ y})  = \sum_{i=1}^K \mathbf a (\theta_i) d_i+  \sigma^2 \mathbf i_L  \in \mathbb C^{L^2 \times 1},
\label{eq:r}
\end{align}
where $\mathrm{vec}(\cdot)$ returns the column-wise vectorization of its matrix argument, $\mathbf a(\theta_i)  \eqdef  \mathbf s(\theta_i)^* \otimes \mathbf s(\theta_i) $, $\mathbf i_L \eqdef \mathrm{vec}(\mathbf I_L) \in \mathbb R^{L^2 \times 1}$,  \textcolor{black}{the superscript `*' denotes complex conjugate,} and $\mathrm{}$ `$\otimes$' is the Kronecker product operator \cite{AGRAHAM}.
By coprime number theory \cite{PP1}, for every $n \in \{-L'+1, -L'+2, \ldots, L'-1\}$ with $L' \eqdef MN+M$, there exists a well-defined set of indices $\mathcal J_{n}  \subset \{ 1, 2, \ldots, L^2\} $, such that
\begin{align}
[\mathbf a(\theta)]_{j} = v(\theta)^n  ~~\forall j\in \mathcal J_n, 
\label{echecka}
\end{align}  
for every $\theta \in (-\frac{\pi}{2}, \frac{\pi}{2}]$. We henceforth consider that $\mathcal J_{n}$ contains all $j$ indices that satisfy \eqref{echecka}.
In view of \eqref{echecka}, a coprime array receiver assembles a linear combining matrix $\mathbf E \in \mathbb R^{L^2 \times 2L'-1}$ and forms a length-$(2L'-1)$ autocorrelation-vector $\mathbf r_{co}$,  each element of which corresponds to a single set  $\mathcal{J}_n$, for every $n \in \{1-L',2-L',\ldots,L'-1\}$, by conducting linear \emph{processing}\footnote{Existing coprime array autocorrelation processing methods in the literature are reviewed in section \ref{SEC: LITERATURE}.} (e.g., $\mathbf E^\top \mathbf r$)  on the autocorrelations in $\mathbf r$.
That is, there exists linear combiner $\mathbf E$ such that 
\begin{align}
\mathbf r_{co}=\mathbf E^\top\mathbf r=\sum_{k=1}^K \mathbf a_{co}(\theta_k)d_k+\sigma^2\mathbf e_{L',2L'-1} ,
\label{eq:rco}
\end{align}
with $\mathbf a_{co}\eqdef\big[v(\theta)^{1-L'},v(\theta)^{2-L'},\ldots,v(\theta)^{L'-1}\big]$ for any $\theta \in (-\frac{\pi}{2},\frac{\pi}{2}]$, and,  for any $p \leq P \in \mathbb N_+$, $\mathbf e_{p,P}$ is the $p$th column of $\mathbf I_{P}$. 
\begin{figure}[!t]
	\includegraphics[width=.98\linewidth]{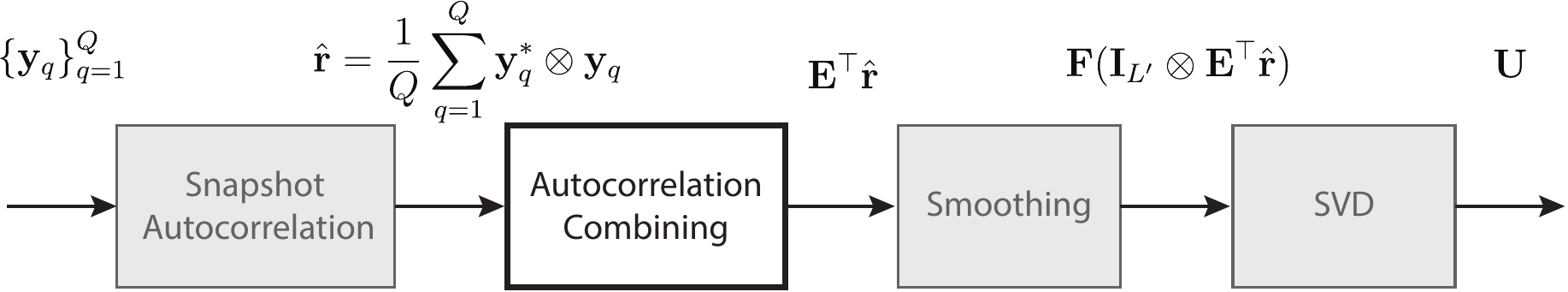}
	\caption{Coprime processing steps: from a collection of samples $\{ \mathbf y_{q} \}_{q=1}^Q$ to the estimated coarray signal-subspace basis $\mathbf U$.}
	\label{fig:steps}
\end{figure}
Thereafter, the receiver applies spatial-smoothing to organize the sampled autocorrelations as the matrix 
\begin{align}
\mathbf Z \eqdef \mathbf F (\mathbf I_{L'} \otimes   \mathbf r_{co}) \in \mathbb C^{L' \times L'},
\label{Zsel}
\end{align}
where $\mathbf F \eqdef [\mathbf F_{1}, \mathbf F_{2}, \ldots, \mathbf F_{L'}]$ and, for every $m \in \{1, 2, \ldots, L'\}$,  $\mathbf F_m \eqdef [\mathbf 0_{L' \times (L'-m)}, \mathbf I_{L'}, \mathbf 0_{L' \times (m-1)}]$.
Importantly,  $\mathbf Z$ coincides with the autocorrelation matrix of a length-$L'$ ULA with antenna elements at locations $\{0,1,\ldots,L'-1\}d$.
That is, 
\begin{align}
\mathbf Z=\mathbf{S}_{\text{co}}\mathrm{diag}(\mathbf d)\mathbf{S}_{\text{co}}^H+\sigma^2\mathbf I_{L'},
\label{eq:Zco}
\end{align}
where it holds that $[\mathbf S_{\text{co}}]_{m,k}=v(\theta_k)^{m-1}$ for every $m \in \{1,2,\ldots,L'\}$ and $k \in \{1,2,\ldots,K\}$.
Standard MUSIC DoA estimation can be applied on $\mathbf Z$.
Let the columns of $\mathbf U \in \mathbb C^{L' \times K}$ be the dominant left-hand singular vectors of $\mathbf Z$, corresponding to its $K$ highest singular values, acquired by means of singular-value-decomposition (SVD).  
Defining $\mathbf v(\theta)=[1, v(\theta), \ldots,v(\theta)^{L'-1}]^\top$, we can accurately decide that $\theta \in (-\frac{\pi}{2},\frac{\pi}{2}]$ belongs in $\Theta\eqdef \{\theta_1, \theta_2, \ldots, \theta_K\}$ if $\left(\mathbf I_{L'}-\mathbf U \mathbf U^H\right)\mathbf v(\theta) =\mathbf 0_{L'}$ is satisfied for some $\theta$.
Equivalently, we can resolve the angles in $\Theta$ by the $K$ (smallest) local minima of the MUSIC spectrum 
\begin{align}
P_{\text{MUSIC}}(\theta)=\Big\|\left(\mathbf I_{L'}-\mathbf U \mathbf U^H\right)\mathbf v(\theta)\Big\|_2^2.
\end{align}
In practice, $\mathbf R_y$ in \eqref{eq:Ry} is unknown to the receiver and sample-average estimated by a collection of $Q$ received-signal snapshots in  $\mathbf Y=\big[\mathbf y_1, \mathbf y_2, \ldots, \mathbf y_Q\big]$ by 
\begin{align}
\hat{\mathbf R}_y=\frac{1}{Q}\sum_{q=1}^{Q}\mathbf y_q \mathbf y_q^H.
\label{eq:Ryhat}
\end{align}
Accordingly, the physical-array autocorrelation-vector $\mathbf r$ in \eqref{eq:r} is estimated by
\begin{align}
\hat{\mathbf r} \eqdef \mathrm{vec}(\hat{\mathbf R}_y) = \frac{1}{Q} \sum_{q=1}^Q \mathbf y_{q}^* \otimes \mathbf y_q.
\label{estR}
\end{align}
The receiver then conducts linear combining on the estimated autocorrelation vector  $\hat{\mathbf r}$ to obtain an estimate of $\mathbf{r}_{co}$, $\hat{\mathbf r}_{co}=\mathbf E^\top \hat{\mathbf r}$. The estimation error $\left\|\mathbf r_{co}-\hat{\mathbf r}_{co}\right\|_2$ depends on how well the linear combiner $\mathbf E$ estimates the nominal physical-array autocorrelations. 
Accordingly, $\mathbf Z$ in \eqref{eq:Zco} is estimated by 
\begin{align}
\hat{\mathbf Z} \eqdef \mathbf F (\mathbf I_{L'} \otimes   \hat{\mathbf r}_{co}) \in \mathbb C^{L' \times L'}.
\label{Zhatsel}
\end{align}
Finally, MUSIC DoA estimation can be applied using the $K$ dominant left-hand singular vectors of $\hat{\mathbf Z}$ instead of those of $\mathbf Z$.
Of course,  in practice, there is an inherent DoA estimation error due to the mismatch between $\mathbf Z$ and $\hat{\mathbf Z}$. 
A schematic illustration of the coprime array processing steps presented above is offered in Fig. \ref{fig:steps}.
In the sequel, we review the most commonly considered autocorrelation combining approaches in the coprime array literature and conduct a formal MSE analysis.
\section{Technical Background on Autocorrelation Combining}
\label{SEC: LITERATURE}
\subsection[Selection Combining]{Selection Combining \cite{PP1}}
The most commonly considered autocorrelation combining method is \emph{selection combining} based on which the receiver selects any single index $j_n \in \mathcal J_n$, for $n \in \{-L'+1,  \ldots, L'-1\}$, and builds the  $L^2 \times (2L'-1)$ \emph{selection}  matrix
\begin{align}
\mathbf E_{sel} \eqdef \left[ \mathbf e_{j_{1-L'}, L^2}, \mathbf e_{j_{2-L'}, L^2}, \ldots, \mathbf e_{j_{L'-1}, L^2} \right],
\label{EQ:E}
\end{align}
by which it processes the autocorrelations in $\mathbf r$, discarding by selection all duplicates (i.e., every entry with index in $\mathcal J_n \setminus j_n$, for every $n$), to form the length-$(2L'-1)$ autocorrelation vector
\begin{align}
{\mathbf r}_{sel} \eqdef \mathbf E_{sel}^\top \mathbf r. 
\label{rsel}
\end{align}
Importantly, when the nominal entries of $\mathbf r$ are known to the receiver, $\mathbf r_{sel}$ coincides with $\mathbf r_{co}$ in \eqref{eq:rco}, thus, applying spatial smoothing on $\mathbf r_{sel}$ yields the exact coarray  autocorrelation matrix
$
\mathbf Z=\mathbf F (\mathbf I_{L'} \otimes  \mathbf r_{sel}).
$  
In contrast, when $\mathbf r$ is unknown to the receiver and estimated by $\hat{\mathbf r}$ in \eqref{estR}, $\mathbf r_{sel}$ in \eqref{rsel}  is estimated by 
\begin{align}
\hat{\mathbf r}_{sel}=\mathbf E_{sel}^\top \hat{\mathbf r}.
\label{eq: rhatsel}
\end{align} 
Accordingly, the coarray autocorrelation matrix is estimated as 
\begin{align}
\hat{\mathbf Z}_{sel}\eqdef\mathbf F (\mathbf I_{L'} \otimes  \hat{\mathbf r}_{sel}) \in \mathbb C^{L' \times L'}.
\end{align}
\subsection[Averaging Combining]{Averaging Combining \cite{CHUNLIU1}}  
Instead of selecting a single index in $\mathcal{J}_n$ by discarding duplicates, averaging combining conducts averaging on all autocorrelation estimates that correspond to $\mathcal{J}_n$ for every $n \in \{1-L',2-L',\ldots, L'-1\}$. 
That is, the receiver assembles the \emph{averaging combining} matrix 
 $\mathbf E_{avg}$, where, for every $i \in \{1, \ldots, 2L'-1\}$, 
 \begin{align}
 [{\mathbf E}_{avg}]_{:,i} \eqdef \frac{1}{|\mathcal J_{i-L'}|} \sum_{j \in \mathcal J_{i-L'}} \mathbf e_{j,L^2}.
 \label{EQ:ECHECK}
 \end{align}
\textcolor{black}{where $|\cdot|$ denotes the cardinality of its argument}. Then, it processes the autocorrelation vector $\mathbf r $ to obtain
 \begin{align}
 \mathbf r_{avg} \eqdef  \mathbf E_{avg}^\top \mathbf r.
 \label{ravg}
 \end{align}
By \eqref{echecka} and the fact that 
$
[\mathbf i_{L}]_{j} $ equals 1, if $j \in \mathcal{J}_0$ and $0$ otherwise, 
it holds that, for any $n \in \{ -L'+1, \ldots, L'-1\}$,  $[\mathbf r]_{j} = \mathbf e_{j, L^2}^\top \mathbf r$ takes a constant value for every $j \in \mathcal J_n$. Thus, 
$\mathbf r_{avg}$ coincides with $\mathbf r_{sel}$ and $\mathbf r_{co}$.
Therefore, similar to the selection combining approach, when $\mathbf r$ is known to the receiver, applying spatial smoothing on $\mathbf r_{avg}$ yields $\mathbf Z=\mathbf F\left(\mathbf I_{L'} \otimes  \mathbf r_{avg}\right).$
In practice, when $\mathbf R_y$ in \eqref{eq:Ry} is estimated by $\hat{\mathbf R}_y $ in \eqref{eq:Ryhat}, $\mathbf r_{avg}$ is estimated by 
\begin{align}
\hat{ \mathbf r}_{avg}\eqdef\mathbf E^\top_{avg}\hat{ \mathbf r},
\label{eq:rhcoavg}
\end{align}
and, accordingly, $\mathbf Z$ is estimated by 
\begin{align}
\hat{\mathbf Z}_{avg}\eqdef\mathbf F (\mathbf I_{L'} \otimes  \hat{\mathbf r}_{avg}) \in \mathbb C^{L' \times L'}.
\end{align}
\subsection[Closed-form MSE Expressions for Selection and Averaging Combining]{Closed-form MSE Expressions for Selection and Averaging Combining \cite{DGC1}}
In general, estimates $\hat{\mathbf r}_{sel}$ and $\hat{ \mathbf r}_{avg}$ diverge from $\mathbf r_{co}$ and attain MSE $e(\hat{\mathbf r}_{sel}) \eqdef \mathbb E\{ \| \mathbf r_{co}  - \hat{\mathbf r}_{sel}  \|_2^2 \}$ and $e(\hat{\mathbf r}_{avg}) \eqdef \mathbb E\{ \| \mathbf r_{co}  - \hat{\mathbf r}_{avg}  \|_2^2 \}$, respectively.
Accordingly, $\hat{\mathbf Z}_{sel}$ and $\hat{\mathbf Z}_{avg}$ diverge from the nominal $\mathbf Z$ and attain MSE $e(\hat{\mathbf Z}_{sel}) \eqdef \mathbb E\{ \| \mathbf Z  - \hat{\mathbf Z}_{sel}  \|_F^2 \}$ and $e(\hat{\mathbf Z}_{avg}) \eqdef \mathbb E\{ \| \mathbf Z  - \hat{\mathbf Z}_{avg}  \|^2_F \}$, respectively.
Closed-form MSE expressions for the errors above were first presented in the form of Lemmas and Propositions (proofs omitted) in \cite{DGC1}. For completeness purposes, we present again the MSE expressions in the form of Lemmas and Propositions, this time, accompanied by formal mathematical proofs.  

For any sample support $Q$, the following Lemma \ref{lem:lemma1} and Lemma \ref{lem:lemma2} express in closed-form the MSE attained by $\hat{\mathbf r}_{sel}$.
\begin{mylemma}
	For any $n \in \left\{-L'+1, -L'+2, \ldots, L'-1\right\}$ and $j \in \mathcal{J}_n$, it holds $e    = \mathbb E\{ |  [\mathbf r]_{j}  - [\hat{\mathbf r}]_{j}  |^2 \} =  \frac{(\mathbf 1_{K}^\top \mathbf d + \sigma^2)^2}{Q}$.
	\hfill $\square$
	\label{lem:lemma1} 
\end{mylemma} 
\begin{mylemma}
$\hat{\mathbf r}_{sel}$ attains MSE 
$e(\hat{\mathbf r}_{sel}) = \mathbb E\left\{ \left\| \mathbf r_{co}  - \hat{\mathbf r}_{sel}  \right\|_2^2 \right\} =\left(2L'-1\right)e.$ 
\hfill $\square$
\label{lem:lemma2} 
\end{mylemma} 
In view of Lemma \ref{lem:lemma2}, the following Proposition follows in a straightforward manner.
\begin{myprop}
$\hat{\mathbf Z}_{sel}$ attains MSE $e(\hat{\mathbf Z}_{sel}) = \mathbb E\left\{ \left\| \mathbf Z - \hat{\mathbf Z}_{sel}  \right\|^2_F \right\} ={L'}^2 e.$
\hfill $\square$
\label{prop:prop1}
\end{myprop}
Expectedly, as the sample-support $Q$ grows asymptotically, $e$, $e(\hat{\mathbf r}_{sel})$, and $e(\hat{\mathbf Z}_{sel})$ converge to zero.

For any sample support $Q$, the following Lemma \ref{lem:lemma3} and Lemma \ref{lem:lemma4} express in closed-form the MSE attained by $\hat{\mathbf r}_{avg}$, where $\dotp\eqdef \mathbf p \otimes \mathbf 1_L$, $\dot{\omega}_{i,j} \eqdef [\dotp]_i-[\dotp]_j$, and
$\mathbf z_{i,j}\eqdef \big[v(\theta_1)^{\dot{\omega}_{i,j}} ~ v(\theta_2)^{\dot{\omega}_{i,j}}~ \ldots~ v(\theta_K)^{\dot{\omega}_{i,j}}\big]^H$.
\begin{mylemma}
For any $n \in \{ -L'+1, \ldots, L'-1 \}$ and $j_n \in \mathcal J_n$, it holds that
\begin{align}
e_{n}  = \mathbb E\left\{ \left|  [\mathbf r]_{j_n}  - \frac{1}{|\mathcal J_{n}|} \sum\limits_{j \in \mathcal J_n} [\hat{\mathbf r}]_{j}  \right|^2 \right\} = \frac{1}{Q} \left( \frac{2\sigma^2 \mathbf 1_{K}^\top \mathbf d +\sigma^4}{|\mathcal J_n|} +  \sum\limits_{i\in \mathcal J_{n}} \sum\limits_{j\in \mathcal J_{n}}\frac{|\mathbf z_{i,j}^H \mathbf d|^2 }{|\mathcal J_{n}|^2} \right).
\end{align}
\hfill $\square$
\label{lem:lemma3}
\end{mylemma}
\begin{mylemma}
 $\hat{\mathbf r}_{avg}$ attains MSE $ e(\hat{\mathbf r}_{avg}) = \mathbb E\left\{ \left\| \mathbf r_{co}  - \hat{\mathbf r}_{avg}  \right\|_2^2 \right\}= \sum_{n=1-L'}^{L'-1}  e_n$.
\hfill$\square$
\label{lem:lemma4} 
\end{mylemma}
By Lemma \ref{lem:lemma4}, the following Proposition naturally follows. 
\begin{myprop}
 $\hat{\mathbf Z}_{avg}$ attains MSE $ e(\hat{\mathbf Z}_{avg}) = \mathbb E\left\{ \left\| \mathbf Z  {-} \hat{\mathbf Z}_{avg}  \right\|^2_F \right\} =\sum\limits_{m=1}^{L'}\sum\limits_{n=1-m}^{L'-m} e_n.$ 
\hfill $\square$
\label{prop:prop2}
\end{myprop}
Similar to selection combining, as the sample-support $Q$ grows asymptotically, $e_n$, $e(\hat{\mathbf r}_{avg})$, and $e(\hat{\mathbf Z}_{avg})$ converge to zero.
Complete proofs for the above Lemmas and Propositions are offered for the first time in the Appendix.

\subsubsection{Remarks on selection and averaging sampling}
In view of the closed-form MSE expressions of selection and averaging combining, few important remarks are drawn.  
\begin{enumerate}[(i)]
	\item In view of Lemma \ref{lem:lemma2} and Lemma \ref{lem:lemma4}, $\hat{\mathbf r}_{sel}$ in \eqref{eq: rhatsel} and $\hat{\mathbf r}_{avg}$ in \eqref{eq:rhcoavg} are both unbiased and consistent estimates of $\mathbf r_{co}$ in \eqref{eq:rco}. 
	That is, as the sample support $Q$ grows asymptotically, both estimates converge provably to $\mathbf r_{co}$. 
	\label{item: item1}
	\item By Propositions \ref{prop:prop1} and \ref{prop:prop2}, it follows that $\hat{ \mathbf Z}_{sel}$ and $\hat{ \mathbf Z}_{avg}$ are unbiased and consistent estimates of $\mathbf Z$.
	Therefore,  MUSIC DoA estimation on $\hat{\mathbf Z}_{sel}$  and $\hat{\mathbf Z}_{avg}$ is expected to perform identically as $Q \rightarrow \infty$ and, thus, very similarly for high values of $Q$. 
	\item 
    For any $i,j$, $| \mathbf z_{i,j}^H \mathbf d|^2 = | \sum_{k=1}^K [\mathbf z_{i,j}]_k^* d_k|^2 \leq | \sum_{k=1}^K | [\mathbf z_{i,j}]_k^* d_k| |^2 = (\mathbf 1_{K}^\top \mathbf d)^2$ and, therefore,  
	\begin{align}
	e-e_n\geq \beta_n\eqdef \frac{|\mathcal J_n|-1}{|\mathcal J_n |Q} (2\sigma^2\mathbf 1_K^\top\mathbf d +\sigma^4)\geq0.
	\label{err1}
	\end{align}
	\item 
	By \eqref{err1}, the MSE difference $e(\hat{\mathbf r}_{sel})-e(\hat{\mathbf r}_{avg})$  is positive,  lower-bounded by ${ \sum_{n=1-L'}^{L'-1} } \beta_n$, and converges to zero as $Q$ increases asymptotically.
	In the realistic case of finite sample support, averaging sampling exhibits superior  estimation performance with respect to MSE, for any value of $Q$.
	\item Similarly, \eqref{err1} implies that  the corresponding MSE difference  $e(\hat{\mathbf Z}_{sel})-e(\hat{\mathbf Z}_{avg})$  is positive,  lower-bounded by $\sum_{m=1}^{L'}{ \sum_{n=1-m}^{L'-m} } \beta_n$, and converges to zero as $Q$ increases asymptotically, which establishes the superiority (in the MSE sense) of averaging combining compared to selection combining in estimating $\mathbf Z$. 
\end{enumerate}
\section{Proposed Minimum Mean-Squared-Error Autocorrelation Combining}
\label{SEC: PROPOSED}
We focus on the autocorrelation combining step of coprime-array processing (see Fig. \ref{fig:steps}) where the receiver applies linear combining matrix $\mathbf E$ to the estimated autocorrelations of the physical array.
Arguably, a preferred receiver will attain consistently (i.e., for any possible configuration of DoAs, $\Theta$) low squared-estimation error $\|\mathbf r_{co}-\mathbf E^\top \hat{ \mathbf r}\|_2^2$.
If this error is exactly equal to zero, then at the fourth step of coprime-array processing, MUSIC will identify the exact DoAs in  $\Theta$. 

In view of the above, in this work, we treat the DoAs in $\Theta$ as independent and identically distributed (i.i.d.) random variables and focus on designing a coprime array receiver equipped with linear combiner $\mathbf E$ such that $\|\mathbf r_{co}-\mathbf E^\top \hat{ \mathbf r}\|_2^2$ is minimized in the mean. 
We assume that, for any $k$,  $\theta_k \in \Theta$ is a random variable with probability distribution $\mathcal D(a,b)$ (e.g., uniform, truncated normal, or, other) where $a$ and $b$ denote the limits of the support of $\mathcal D$ and seek the minimum mean-squared-error combining matrix $\mathbf E$ which minimizes $\mathbb E \left\{\|\mathbf r_{co}-\mathbf E^\top \hat{ \mathbf r}\|_2^2\right\}$. In fact, probability distributions of angular variables is not a new concept. 
Angular variables have been modeled by the {\it{von Mises}} Probability Density Function (PDF) which can include (or, nearly approximate) standard distributions such as Uniform, Gaussian, and wrapped Gaussian, to name a few,  by tuning a parameter in the PDF expression \cite{randomvDoA0,randomvDoA1}.
Angular distributions have also been considered for Bayesian-based beamforming in \cite{randomvDoA2}. More recently, angular distributions were also considered for Bayesian-based beamforming for millimeter wave channel tracking \cite{randomvDoA3}.

In the most general case and in lieu of any pertinent prior information at the receiver, the DoAs in $\Theta$ can, for instance, be assumed to be the uniformly distributed in $(-\frac{\pi}{2},\frac{\pi}{2}]$--i.e., $\mathcal D(a,b)\equiv\mathcal U(-\frac{\pi}{2},\frac{\pi}{2})$. In the sequel, we derive the minimum mean-squarer-error combining matrix for any continuous probability distribution $\mathcal D(a,b)$ with $-\frac{\pi}{2}<a<b\leq \frac{\pi}{2}$.

First, we introduce new notation on the problem statement and formulate the MSE minimization problem. 
By defining
\begin{align}
\mathbf A\eqdef\left[\mathbf S \ \mathrm{diag}([\sqrt{d_1},\sqrt{d_2},\ldots,\sqrt{d_K}]), \sigma \mathbf I_L\right] \in \mathbb C^{L\times K+L},
\end{align}
the autocorrelation matrix $\mathbf R_y$ in \eqref{eq:Ry} is factorized as $\mathbf R_y=\mathbf A \mathbf A^H.$
Accordingly, $\mathbf r$ in \eqref{eq:r} can be  expressed as 
$ \mathbf r=\left(\mathbf A^*\otimes \mathbf A\right)\mathrm{vec}\left(\mathbf I_{K+L}\right)$.
Moreover, we define 
\begin{align}
\mathbf V\eqdef \mathbf A^* \otimes \mathbf A \in  \mathbb C^{L^2 \times \left(K+L\right)^2},
\label{eq: Vmatrix}
\end{align} 
and $\mathbf i\eqdef\mathrm{vec}\left(\mathbf I_{K+L}\right) \in \mathbb R^{(K+L)^2}$, where $\mathbf I_{K+L}$ is the $(K+L)$-size identity matrix. 
Then, $\mathbf r$ takes the form
\begin{align}
\mathbf r =\mathbf V \mathbf i.
\label{eq:rVi} 
\end{align}
It follows that $\mathbf r_{co}$ (or, $\mathbf r_{sel}$ and $\mathbf r_{avg}$ in \eqref{rsel} and \eqref{ravg}, respectively) can be  expressed as $\mathbf E_{sel}^\top \mathbf V \mathbf i=\mathbf E_{avg}^\top \mathbf V \mathbf i$.
Next, we observe that for every $q \in \{1,2,\ldots,Q\}$, there exists a vector $\mathbf x_q \sim \mathcal{CN}(\mathbf 0_{K+L}, \mathbf I_{K+L})$, pertinent to $\mathbf y_q$, such that 
$
\mathbf y_q=\mathbf A\mathbf x_q.
$
From the signal model, $\mathbf x_q$ is statistically independent from $\mathbf x_p$ for any pair $(p,q) \in \{1,2,\ldots,Q\}^2$  with $p\neq q$.
Then, the physical-array autocorrelation matrix estimate $\hat{\mathbf R}_y$ in \eqref{eq:Ryhat}, is expressed as $\hat{\mathbf R}_y=\mathbf A \mathbf W \mathbf A^H$, where $\mathbf W\eqdef \frac{1}{Q}\sum_{q=1}^{Q}\mathbf x_q \mathbf x_q^H$. 
Moreover, by defining
$
\mathbf w \eqdef \mathrm{vec}\left(\mathbf W\right)=\frac{1}{Q}\sum_{q=1}^{Q}\mathbf x_q^*\otimes \mathbf x_q,
$ 
the estimate $\hat{ \mathbf r}$ in \eqref{estR} takes the form
\begin{align}
\hat{\mathbf r}=\mathbf V\mathbf w.
\label{eq:rVw}
\end{align}
In view of \eqref{eq:rVi} and \eqref{eq:rVw}, we propose to design a MMSE linear combiner, $\mathbf E_{\text{MMSE}}$, by formulating and solving the MSE-minimization problem 
\begin{align}
\argmin_{\mathbf E \in \mathbb C^{L^2 \times 2L'-1}} \underset{\Theta, \mathbf w}{\mathbb E} \left\lbrace\left\|\mathbf E^H\mathbf V \mathbf w-\mathbf E_{sel}^\top\mathbf V \mathbf i\right\|_2^2\right\rbrace.
\label{eq:newprob}
\end{align}
Of course, if we replace $\mathbf E_{sel}$ by $\mathbf E_{avg}$ in \eqref{eq:newprob}, the resulting problem will be equivalent to \eqref{eq:newprob}.
In the sequel, we show that a closed-form solution for \eqref{eq:newprob} exists for any finite value of sample support $Q$ and present a step-by-step solution.

We commence our solution by defining $\mathbf G\eqdef \mathbf V\mathbf w \mathbf w^H \mathbf V^H \in \mathbb C^{L^2 \times L^2}$ and $\mathbf H\eqdef\mathbf V \mathbf w \mathbf i^H \mathbf V^H \in \mathbb C^{L^2 \times L^2}$.
Then, the problem in \eqref{eq:newprob} simplifies to 
\begin{align}
\argmin_{\mathbf E \in \mathbb C^{L^2 \times 2L'-1}} \underset{\Theta, \mathbf w}{\mathbb E} \Big\{
\mathrm{Tr}\left(\mathbf E^H\mathbf G \mathbf E\right)-2\Re\left\lbrace\mathrm{Tr}\left(\mathbf E^H\mathbf H\mathbf E_{sel}\right)\right\rbrace\Big\},
\label{eq:newprob2}
\end{align}
where $\Re\{\cdot\}$ extracts the real part of its argument and $\mathrm{Tr}(\cdot)$  returns the sum of the diagonal entries of its argument. 
Furthermore, we define $\mathbf G_{\mathbb E}\eqdef \mathbb E_{\Theta, \mathbf w}\{\mathbf G\}$ and $\mathbf H_{\mathbb E}\eqdef \mathbb E_{\Theta, \mathbf w}\{\mathbf H\}$. Then,  $\eqref{eq:newprob2}$ takes the equivalent form
\begin{align}
\argmin_{\mathbf E \in \mathbb C^{L^2 \times 2L'-1}}  \Big\{
\mathrm{Tr}\left(\mathbf E^H\mathbf G_{\mathbb E} \mathbf E\right)-2\Re\left\lbrace\mathrm{Tr}\left(\mathbf E^H\mathbf H_{\mathbb E}\mathbf E_{sel}\right)\right\rbrace\Big\}.
\label{eq:newprob3}
\end{align}  
Next, we focus on deriving closed-form expressions for $\mathbf G_{\mathbb E}$ and $\mathbf H_{\mathbb E}$ that will allow us to solve \eqref{eq:newprob3} and obtain the minimum-MSE linear combiner $\mathbf E_{\text{MMSE}}$.
At the core of our developments lies the observation that, for any $\theta \sim \mathcal{D}(a,b)$ with PDF $f(\theta)$ and scalar $x \in \mathbb R$, it holds that
\begin{align}
\underset{\theta}{\mathbb E}\{v(\theta)^x\}&=\int_{a}^{b}f(\theta)\mathrm {exp}\left(-j x\frac{2\pi f_c}{c}\sin \theta\right)d\theta \eqdef \mathcal I(x).
\label{general}
\end{align}
\begin{figure}[!t]
	\centering
	\includegraphics[width=.5\linewidth]{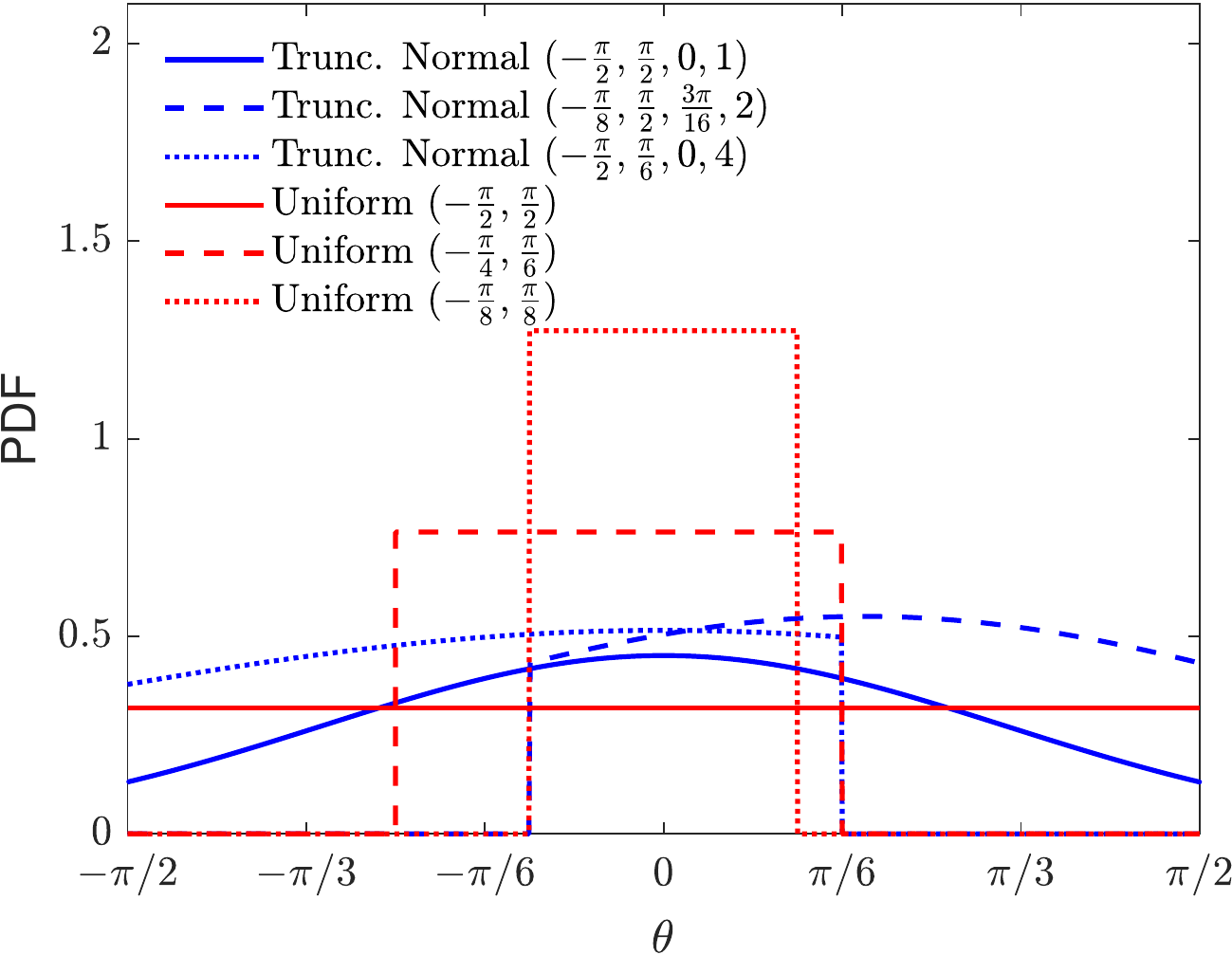}
	\caption{Probability density function $f(\theta)$ for different distributions and support sets.}
	\label{fig:distributions}
\end{figure}
The integral $\mathcal I(x)$ can be approximated within some numerical error tolerance with numerically efficient vectorized methods \cite{integralVectorized}. 
In Fig. \ref{fig:distributions}, we offer visual illustration examples of $f(\theta)$ when $\theta$ follows a uniform distribution in $(a,b)$, i.e., $\theta \sim \mathcal{U}(a,b)$, or, a truncated normal distribution in $(a,b)$ with mean $\mu$ and variance $\sigma^2$, i.e., $\theta\sim\mathcal {TN}(a,b,\mu,\sigma^2)$. More specifically, 
\begin{align}
f(\theta)=\begin{cases}\frac{\frac{1}{\sigma \sqrt{2\pi}}\mathrm{exp}\left(-\frac{1}{2}\left(\frac{\theta-\mu}{\sigma}\right)^2\right)}{\frac{1}{2}\left(\mathrm{erf}\left(\frac{b-\mu}{\sigma\sqrt{2}}\right)-\mathrm{erf}\left(\frac{a-\mu}{\sigma\sqrt{2}}\right)\right)},\quad \theta\sim\mathcal{TN}(a,b,\mu,\sigma^2),\\\frac{1}{b-a} ,\quad \quad\quad\quad\quad\quad\quad\quad\quad\theta\sim\mathcal U(a,b),\end{cases}
\end{align}
where $\mathrm{erf}(\cdot)$ denotes the Error Function \cite{erf}.
In the special case that $\mathcal D(a,b)\equiv \mathcal U(-\frac{\pi}{2}, \frac{\pi}{2})$, $\mathcal I(x)$ coincides with $J_0(x\frac{2 \pi f_c}{c})$: the $0$-th order Bessel function of the first kind \cite{BESSEL} for which there exist look-up tables. 
Next, we define the indicator function $\delta(x)$ which equals $1$ if $x=0$ and assumes a value of zero otherwise.
We provide in the following Lemma the statistics of the random variable $\mathbf w$ which appears in the closed-form expressions of $\mathbf G_{\mathbb{E}}$ and $\mathbf H_{\mathbb{E}}$. 
\begin{mylemma}
The first- and second-order statistics of the random variable $\mathbf w$ are given by
$
	\mathbb E_{\mathbf w}\left\{\mathbf w\right\}=\mathbf i \in \mathbb R^{(K+L)^2 } $  and $ \mathbb E_{\mathbf w}\left\{\mathbf w \mathbf w^H\right\}=\mathbf i\mathbf i^\top +\frac{1}{Q}\mathbf I_{(K+L)^2} \in \mathbb R^{(K+L)^2 \times (K+L)^2},
	$
	respectively. 
	\hfill $\square$
	\label{lem:w_statistics}
\end{mylemma}
A  proof for Lemma \ref{lem:w_statistics} is offered in the Appendix. 
Next, we define $\ddotp \eqdef \mathbf 1_L \otimes \mathbf p$ and $\omega_i\eqdef [\dotp]_i-[\ddotp]_i$. In view of Lemma \ref{lem:w_statistics}, we present an entry-wise closed-form expression for $\mathbf H_{\mathbb E}$ in the following Lemma. 
\begin{mylemma}
For any $(i,m) \in \{1,2,\ldots,L^2\}^2$,
	\begin{align}
	[\mathbf H_{\mathbb E}]_{i,m}&=\left\|\mathbf d\right\|_2^2\mathcal{I}\left(\omega_i-\omega_m\right)+\sigma^4\delta(\omega_i)\delta(\omega_m)
	+\mathcal{I}(\omega_i)\mathcal{I}(-\omega_m)\left((\mathbf 1_K^\top\mathbf d)^2
	-\left\|\mathbf d\right\|_2^2\right)
	\\&+\sigma^2\left(\mathbf 1_K^\top\mathbf d\right)\Big(\delta(\omega_i)\mathcal{I}(-\omega_m)+\mathcal{I}(\omega_i)\delta(-\omega_m)\Big).
	\label{Jeim}
	\end{align}
    \hfill $\square$
	\label{lemmaHe}
\end{mylemma}
A complete proof for Lemma \ref{lemmaHe} is also provided in the Appendix.
Hereafter, we focus on deriving a closed-form expression for $\mathbf G_{\mathbb E}$.
First, we define $\tilde{\mathbf V}\eqdef \mathbf V\mathbf V^H$ whose expectation, $\tilde{\mathbf V}_{\mathbb E} \eqdef \underset{\Theta}{\mathbb E}\{ \tilde{\mathbf V}\}$, appears in $\mathbf G_{\mathbb E}$.
We observe that each entry of $\tilde{\mathbf V}$ can be expressed as a linear combination of the entries of $\mathbf V$. 
That is, for any $(i,m) \in \{1,2,\ldots,L^2\}^2$ and $j \in \{1,2,\ldots,(K+L)^2\}$, it holds  
\begin{align}
[\tilde{\mathbf V}]_{i,m}=\sum_{j=1}^{(K+L)^2}[\mathbf V]_{i,j} [\mathbf V^*]_{m,j}.
\label{eq: vtilde}
\end{align}

An entry-wise closed-form for $\mathbf V$, in terms of $\Theta$, is offered in Table \ref{table:Vmatrix}. 
Then, for any triplet $(i,m,j)$ such that $(i,m) \in \{1,2,\ldots,L^2\}^2$ and $j \in \{1,2,\ldots,(K+L)^2\}$, we derive  a closed-form expression for
\begin{align}
\gamma_j^{(i,m)}\eqdef [\mathbf V]_{i,j} [\mathbf V^*]_{m,j},
\label{gamma_ijm}
\end{align}
in Table \ref{table: gamma}.
\begin{table}[!t]
	\centering
	\caption{Entry-wise closed-form expression for matrix $\mathbf V$  defined in \eqref{eq: Vmatrix}.}
	\label{table:Vmatrix}
	\resizebox{.85\linewidth}{!}{%
		\begin{tabular}{ll}
			\toprule
		    $[\mathbf V]_{i,j}$ & Condition on (i,j) \\ \midrule\midrule
			$\sqrt{[\mathbf d]_{\ddot{\mathbf u}_j}[\mathbf d]_{\dot{\mathbf u}_j}}v(\theta_{[\ddot{\mathbf u}_j]})^{[\ddot{\mathbf p}_i]}v(\theta_{[\dot{\mathbf u}]_j})^{-[\dot{\mathbf p}]_i}$ & $[\ddot{\mathbf u}]_j,[\dot{\mathbf u}]_j \leq K$ \\
			$\sigma\sqrt{[\mathbf d]_{\dot{\mathbf u}_j}}v(\theta_{[\dot{\mathbf u}]_j})^{-[\dot{\mathbf p}]_i}$ & $[\dot{\mathbf u}]_j \leq K \text{ and }[\ddot{\mathbf u}]_j= [\ddot{\mathbf v}]_i +K$ \\
			$\sigma\sqrt{[\mathbf d]_{\ddot{\mathbf u}_j}}v(\theta_{[\ddot{\mathbf u}_j]})^{[\ddot{\mathbf p}_i]}$ & $[\ddot{\mathbf u}]_j \leq K \text{ and }[\dot{\mathbf u}]_j= [\dot{\mathbf v}]_i +K$ \\
			$\sigma^2$ & $[\dot{\mathbf u}]_j= [\dot{\mathbf v}]_i +K \text{ and }[\ddot{\mathbf u}]_j= [\ddot{\mathbf v}]_i +K$ \\
			$0$ & otherwise \\
			\hline
		\end{tabular}%
	}
	\resizebox{.87\linewidth}{!}{%
    \footnotesize{
		\begin{tabular}{l}
			\hline
			Auxiliary variables used in  the above conditions/expressions: \\
		 $\mathbf s_x=[1,2,\ldots,x]^\top$,  $\dot{\mathbf u}=\mathbf 1_{K+L}\otimes \mathbf s_{K+L}$,   $\ddot{\mathbf u}=\mathbf s_{K+L}\otimes \mathbf 1_{K+L}$, $\dot{\mathbf v}=\mathbf 1_{L}\otimes \mathbf s_L$, $\ddot{\mathbf v}=\mathbf s_L\otimes \mathbf 1_{L}$.\\ 
			\bottomrule
		\end{tabular}%
	}
	}
\end{table}

Accordingly, $\mathbb E_{\Theta} \{\gamma_j^{(i,m)}\}$ is offered in Table \ref{table: Exp_gamma}, based on which, the $(i,m)$-th entry of $\tilde{\mathbf V}_{\mathbb E}$ is computed as
\begin{align}
[\tilde{\mathbf V}_{\mathbb E}]_{i,m}=\sum_{j=1}^{(K+L)^2} \mathbb E_{\Theta} \{\gamma_j^{(i,m)}\}.
\end{align}

The closed-form expression for matrix $\mathbf G_{\mathbb E}$ is provided in the following Lemma. 
\begin{mylemma}
Matrix $\mathbf G_{\mathbb E}$ is given by  
	$
	\mathbf G_{\mathbb E}=\mathbf H_{\mathbb E}+\frac{1}{Q}\tilde{\mathbf V}_{\mathbb E}.
	$
	\label{lemmaGe}
	\hfill $\square$
\end{mylemma}
A proof for Lemma \ref{lemmaGe} is offered in the Appendix.
We differentiate \eqref{eq:newprob3} with respect to $\mathbf E$, set its derivative to zero, and obtain
\begin{align}
\left(\mathbf H_{\mathbb E}+\frac{1}{Q}\tilde{\mathbf V}_{\mathbb E}\right)\mathbf E_{\text{MMSE}}=\mathbf  H_{\mathbb E} \mathbf E_{sel}.
\label{eq:newprob4}
\end{align}
We observe that \eqref{eq:newprob4} is, in practice, a collection of $(2L'-1)$ systems of linear equations. Let $\mathbf E_{\text{MMSE}}=[\mathbf e_1, \ldots, \mathbf e_{2L'-1}]$ and $\mathbf c_i=[\mathbf H_{\mathbb E} \mathbf E_{sel}]_{:,i}~\forall i \in  \{1,2,\ldots, 2L'-1\}$. Solving \eqref{eq:newprob4} is equivalent to solving, for every $i$,  
\begin{align}
\mathbf G_{\mathbb E}\mathbf e_i =\mathbf c_i.
\label{eq:ci}
\end{align}
\begin{table}[!t]
	\centering
	\caption{Closed-form expression for $\gamma_j^{(i,m)}$ defined in \eqref{gamma_ijm} .}
	\label{table: gamma}
	\resizebox{.85\linewidth}{!}{%
		\begin{tabular}{ll}
			\toprule
			$\gamma_j^{(i,m)}$ & Condition on $(i,m,j)$   \\ \midrule \midrule
			$[\mathbf d]_{\ddot{\mathbf u}_j}[\mathbf d]_{\dot{\mathbf u}_j}v(\theta_{[\ddot{\mathbf u}_j]})^{[\ddot{\mathbf p}_i]-[\ddot{\mathbf p}_m]}v(\theta_{[\dot{\mathbf u}]_j})^{[\dot{\mathbf p}]_m-[\dot{\mathbf p}]_i}$ & $[\dot{\mathbf u}]_{j}{=}[\ddot{\mathbf u}]_{j}\leq K,$  \\
			$[\mathbf d]_{\ddot{\mathbf u}_j}[\mathbf d]_{\dot{\mathbf u}_j}v(\theta_{[\ddot{\mathbf u}_j]})^{[\ddot{\mathbf p}_i]-[\ddot{\mathbf p}_m]}v(\theta_{[\dot{\mathbf u}]_j})^{[\dot{\mathbf p}]_m-[\dot{\mathbf p}]_i}$ & $[\dot{\mathbf u}]_{j},[\ddot{\mathbf u}]_{j}\leq K ;~ [\ddot{\mathbf u}]_{j}\neq [\dot{\mathbf u}]_{j}$    \\ 
			$\sigma^2[\mathbf d]_{\dot{\mathbf u}_j}v(\theta_{[\dot{\mathbf u}]_j})^{[\dot{\mathbf p}]_m-[\dot{\mathbf p}]_i}$ & $[\dot{\mathbf u}]_{j} \leq K ;~ [\ddot{\mathbf u}]_{j}{-}K{=}[\ddot{\mathbf v}]_{i}{=}[\ddot{\mathbf v}]_{m}$   \\ 
			$\sigma^2[\mathbf d]_{\ddot{\mathbf u}_j}v(\theta_{[\ddot{\mathbf u}_j]})^{[\ddot{\mathbf p}]_i-[\ddot{\mathbf p}]_m}$ & $[\ddot{\mathbf u}]_{j} \leq K ;~ [\dot{\mathbf u}]_{j}{-}K{=}[\dot{\mathbf v}]_{i}{=}[\dot{\mathbf v}]_{m}$  \\ 
			{$\sigma^4$} & $[\dot{\mathbf u}]_{j}{-}K{=}[\dot{\mathbf v}]_{i}{=}[\dot{\mathbf v}]_{m} ; ~[\ddot{\mathbf u}]_{j}{-}K{=}[\ddot{\mathbf v}]_{i}=[\ddot{\mathbf v}]_{m}$ \\ 
	     $	0$ & otherwise \\ \midrule
		\end{tabular}%
		}
		\resizebox{.87\linewidth}{!}{%
			\footnotesize{
				\begin{tabular}{l}
					\hline
					Auxiliary variables used in  the above conditions/expressions: \\
					$\dotp= \mathbf p \otimes \mathbf 1_L$, $\ddotp=\mathbf 1_L \otimes \mathbf p$, $\mathbf s_x=[1,2,\ldots,x]^\top$,  $\dot{\mathbf u}=\mathbf 1_{K+L}\otimes \mathbf s_{K+L}$,   $\ddot{\mathbf u}=\mathbf s_{K+L}\otimes \mathbf 1_{K+L}$, \\$\dot{\mathbf v}=\mathbf 1_{L}\otimes \mathbf s_L$, $\ddot{\mathbf v}=\mathbf s_L\otimes \mathbf 1_{L}$.\\ 
					\bottomrule
				\end{tabular}%
			}
		}
\end{table}

\begin{table}[!t]
	\centering
	\caption{closed-form expression for $\mathbb E_{\Theta} \{\gamma_j^{(i,m)}\}$.}
	\label{table: Exp_gamma}
	\resizebox{.85\linewidth}{!}{%
		\begin{tabular}{@{}ll@{}}
			\toprule
			$\mathbb E_{\Theta}\{\gamma_j^{(i,m)}\}$ & Condition on $(i,m,j)$ \\ \midrule
	     	$[\mathbf d]_{[\ddot{\mathbf u}]_{j}}[\mathbf d]_{[\dot{\mathbf u}]_{j}}\mathcal{I}(\ddot{\omega}_{m,i}+\dot{\omega}_{i,m})$ & $[\dot{\mathbf u}]_{j}{=}[\ddot{\mathbf u}]_{j}\leq K,$ \\
			$[\mathbf d]_{[\ddot{\mathbf u}]_{j}}[\mathbf d]_{[\dot{\mathbf u}]_{j}}\mathcal{I}(\ddot{\omega}_{m,i})\mathcal{I}(\dot{\omega}_{i,m})$ & $[\dot{\mathbf u}]_{j},[\ddot{\mathbf u}]_{j}\leq K ;~ [\ddot{\mathbf u}]_{j}\neq [\dot{\mathbf u}]_{j}$ \\
			$\sigma^2 [\mathbf d]_{ [\dot{\mathbf u}]_{j}}\mathcal{I}(\ddot{\omega}_{m,i})$ & $[\dot{\mathbf u}]_{j} \leq K ;~ [\ddot{\mathbf u}]_{j}{-}K{=}[\ddot{\mathbf v}]_{i}{=}[\ddot{\mathbf v}]_{m}$ \\
			$\sigma^2 [\mathbf d]_{ [\ddot{\mathbf u}]_{j}}\mathcal{I}(\dot{\omega}_{i,m})$ & $[\ddot{\mathbf u}]_{j} \leq K ;~ [\dot{\mathbf u}]_{j}{-}K{=}[\dot{\mathbf v}]_{i}{=}[\dot{\mathbf v}]_{m}$ \\
            $\sigma^4$ & $[\dot{\mathbf u}]_{j}{-}K{=}[\dot{\mathbf v}]_{i}{=}[\dot{\mathbf v}]_{m} ; ~[\ddot{\mathbf u}]_{j}{-}K{=}[\ddot{\mathbf v}]_{i}=[\ddot{\mathbf v}]_{m}$ \\
            $0$ & otherwise\\  \midrule
		\end{tabular}%
		}
		\resizebox{.87\linewidth}{!}{%
			\footnotesize{
				\begin{tabular}{l}
					\hline
					Auxiliary variables used in the above conditions/expressions: \\
					$\dotp= \mathbf p \otimes \mathbf 1_L$, $\ddotp=\mathbf 1_L \otimes \mathbf p$, 	$\dot{\omega}_{i,j}=[\dot{\mathbf p}]_i-[\dot{\mathbf p}]_j$,
					$\ddot{\omega}_{i,j}=[\ddot{\mathbf p}]_i-[\ddot{\mathbf p}]_j$, 
					$\mathbf s_x=[1,2,\ldots,x]^\top$,\\ 
					$\dot{\mathbf u}=\mathbf 1_{K+L}\otimes \mathbf s_{K+L}$,   
					$\ddot{\mathbf u}=\mathbf s_{K+L}\otimes \mathbf 1_{K+L}$, 
					$\dot{\mathbf v}=\mathbf 1_{L}\otimes \mathbf s_L$, $\ddot{\mathbf v}=\mathbf s_L\otimes \mathbf 1_{L}$.\\ 
					\bottomrule
				\end{tabular}%
			}
		}
\end{table}
For any $i$ such that $\mathbf c_i \in \mathrm{span}(\mathbf G_{\mathbb E})$, \eqref{eq:ci} has at least one exact solution $\mathbf e_i=\mathbf V\mathbf \Sigma^{-1}\mathbf U^H\mathbf c_i +\mathbf b_i$, where $\mathbf G_{\mathbb E}$ admits SVD $\mathbf U_{L^2 \times \rho} \mathbf \Sigma_{\rho \times \rho}\mathbf V_{\rho \times L^2}^H$, $\rho=\mathrm{rank}(\mathbf G_{\mathbb E})$, and $\mathbf b_i$ is an arbitrary vector in the nullspace of $\mathbf G_{\mathbb E}$ which is denoted by $\mathcal N(\mathbf G_{\mathbb E})$. In the special case that $\rho=L^2$, that is, $\mathbf G_{\mathbb E}$ has full-rank, then $\mathcal N(\mathbf G_{\mathbb E})=\mathbf 0_{L^2}$ and there exists a unique solution $\mathbf e_i=\mathbf V\mathbf \Sigma^{-1}\mathbf U^H\mathbf c_i$. If, on the other hand, $\exists i$ such that $\mathbf c_i \notin \mathrm{span}(\mathbf G_{\mathbb E})$, then \eqref{eq:ci} does not have an exact solution and a Least Squares (LS) approach can be followed by solving $\min_{\mathbf e_i}\|\mathbf G_{\mathbb E}\mathbf e_i-\mathbf c_i\|_2^2$. Interestingly, it is easy to show that the LS solution is the same as before, i.e., $\mathbf e_i=\mathbf V\mathbf \Sigma^{-1}\mathbf U^H\mathbf c_i +\mathbf b_i$, where $\mathbf b_i \in \mathcal N(\mathbf G_{\mathbb E})$.  In every case, each column of $\mathbf E_{\text{MMSE}}$ can be computed in closed-form as 
\begin{align}\mathbf e_i=\mathbf V\mathbf \Sigma^{-1}\mathbf U^H\mathbf c_i+\mathbf b_i, \quad \mathbf b_i \in \mathcal N(\mathbf G_{\mathbb E}).\end{align}

In view of the above, we propose to process the autocorrelations in $\hat{\mathbf{r}}$ by the linear combiner $\mathbf E_{\text{MMSE}}$ to obtain the MMSE estimate of $\mathbf r_{co}$,
\begin{align}
\hat{\mathbf r}_{\text{MMSE}}\eqdef\mathbf E_{\text{MMSE}}^\top\hat{\mathbf r}.
\end{align} 
In turn, we propose to minimum-MSE estimate $\mathbf Z$ in \eqref{Zsel} by  
\begin{align}
\hat{\mathbf Z}_{\text{MMSE}}\eqdef\mathbf F (\mathbf I_{L'} \otimes  \hat{\mathbf r}_{\text{MMSE}}) \in \mathbb C^{L' \times L'}.
\label{Zmmse}
\end{align} 
The proposed linear combiner $\mathbf E_{\text{MMSE}}$ depends on $\mathbf H_{\mathbb E}$ and $\tilde{\mathbf V}_{\mathbb E}$ which, in turn, depend on the powers $d_1, d_2, \ldots,d_K,\sigma^2$ associated to the source DoAs $\theta_1, \theta_2, \ldots, \theta_K$ and noise, respectively.
In general, the receiver has no prior knowledge of these powers. Thus, it cannot compute $\mathbf E_{\text{MMSE}}$ directly. 
This drawback can be addressed with one of the following ways. 
First, similar to the DoAs $\{\theta_k\}_{k=1}^K$, we can assume that the powers $\{d_k\}_{k=1}^K$ and $\sigma^2$ are random variables that are statistically independent from each other and of every DoA, following a specific probability distribution (e.g., uniform). In this case, the proposed linear combiner $\mathbf E_{\text{MMSE}}$ will cease to depend on $\{d_k\}_{k=1}^K, \sigma^2$ and can be deterministically formed at the receiver, independently from the DoAs $\{\theta_k\}_{k=1}^K$, their respective powers $\{d_k\}_{k=1}^K$, and noise power $\sigma^2$.
Alternatively, we can estimate the powers associated with the DoAs $\{\theta_k\}_{k=1}^K$ and noise power in order to compute the linear combiner $\mathbf E_{\text{MMSE}}$.
A simple manner in which to estimate $\{d_k\}_{k=1}^K$ is the Minimum Variance Distortion-less Response (MVDR) spectrum, also known as Capon spectrum \cite{HIRAKAWA,CHEN,DELAMARE,DELAMARE2}, or, its robust version \cite{LARDIES}.
More specifically, we can estimate the DoAs $\{\theta_k\}_{k=1}^K$ by applying MUSIC on $\hat{\mathbf Z}_{avg}$,  obtaining $\{\hat{\theta}_{k}\}_{k=1}^{K}$. Then, the power $d_k$ of the source with DoA $\theta_k$ can be estimated by\footnote{Recall that $\mathbf v(\theta)$ is defined in Section \ref{SEC: SIGNAL}, for any $\theta \in (-\frac{\pi}{2},\frac{\pi}{2}]$.} $\hat{d}_k=(\mathbf v^H(\hat{\theta}_{k}) \hat{\mathbf Z}_{avg}^{-1} \mathbf v(\hat{\theta}_{k}))^{-1}$, for every $k=1,2,\ldots,K$.
The noise power can be estimated by the square root of the smallest eigenvalue of $\hat{\mathbf Z}_{avg}\hat{\mathbf Z}_{avg}^H$, yielding $\hat{\sigma}^2$.
The estimates $\{\hat{d}_k\}_{k=1}^K$ and $\hat{\sigma}^2$ can then be used for approximating $\mathbf E_{\text{MMSE}}$. Of course, there exist more sophisticated algorithms for estimating the source powers (e.g., \cite{SHEN} estimated the powers $\{d_k\}_{k=1}^K$ based on a Vandermonde decomposition of $\sqrt{\frac{1}{L'}\hat{\mathbf Z}_{sel}\hat{\mathbf Z}_{sel}^H}$).  Interestingly enough, however, the receiver needs not know the exact powers $\{d_k\}_{k=1}^K,\sigma^2$.
In fact, knowledge of the power ratios suffices to form the exact MMSE combiner. 
Without loss of generality, we can assume the ratios $\frac{d_2}{d_1}, \frac{d_3}{d_1},\ldots,\frac{d_K}{d_1},\frac{\sigma^2}{d_1}$ to be known and then, $\mathbf E_{\text{MMSE}}$ can be computed exactly. It is easy to see that substituting $d_2,d_3,\ldots,d_K,\sigma^2$ in the closed-form expressions of the columns of $\mathbf E_{\text{MMSE}}$ by the corresponding ratios above does not result in any change.

\section{Numerical Studies}

 \label{SEC:NUMERICAL}
We consider comprime naturals $(M,N)$ with $M<N$ and form an $L$-element physical coprime array, which yields a length-($L'=MN+M$) ULA as the virtual coarray. 
Signals from $K$ sources impinge on the array with equal transmit power $d_1=d_2=...=d_K=\alpha^2= 10$ dB.
The noise variance is fixed at $\sigma^2=0$ dB. Accordingly, the signal-to-noise-ratio (SNR) is equal to $10$ dB for every DoA signal source.

\begin{figure}[!t]
	\centering
	\includegraphics[width=\linewidth]{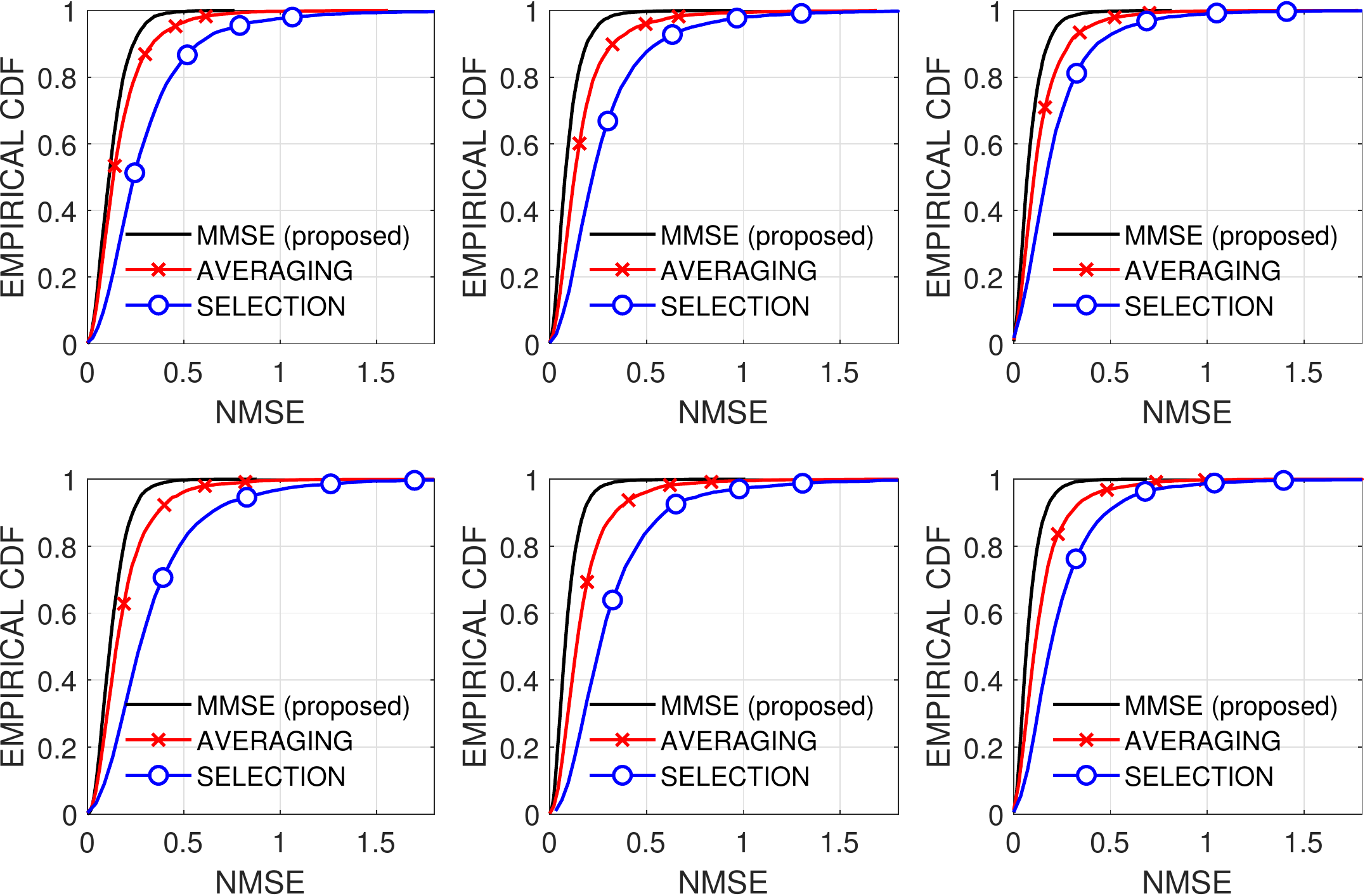}
	\caption{Empirical CDF of the MSE in estimating $\mathbf Z_{}$ for $(M,N)=(2,3)$, SNR$=10$dB, $Q=10$, $K=5$ (top), and $K=7$ (bottom). $\forall k, \theta_k\sim$  $\mathcal U(-\frac{\pi}{2},\frac{\pi}{2})$ (left), $\mathcal U(-\frac{\pi}{4},\frac{\pi}{6})$ (center), $\mathcal{TN}(-\frac{\pi}{8},\frac{\pi}{8},0,1)$ (right).}
	\label{fig:cdf1}
\end{figure}

We commence our studies by computing the empirical Cumulative Distribution Function (CDF) of the Normalized-MSE (NMSE) in estimating $\mathbf Z$ for a given DoA collection $\Theta=\{\theta_1,\ldots, \theta_K\}$ such that the DoAs in $\Theta$ are i.i.d., i.e., $\theta_k\sim\mathcal D(a,b) \ \forall k$. More specifically, we consider fixed sample-support $Q=10$ and for each estimate $\hat{\mathbf Z} \in \{\hat{ \mathbf Z}_{sel}$, $\hat{ \mathbf Z}_{avg},\hat{ \mathbf Z}_{\text{MMSE}}\}$, we compute the estimation error
\begin{align}
\text{MNSE}={\left\|\mathbf{Z}-\hat{\mathbf{Z}}_{}\right\|_F^2}{\left\|\mathbf Z_{}\right\|_F^{-2}}.
\end{align}
We repeat this process over 4000 statistically independent realizations of $\Theta$ and noise and collect 4000 NMSE measurements. Based on these NMSE measurements, we plot in Fig. \ref{fig:cdf1} the empirical CDF of the NMSE in estimating $\mathbf Z$ for fixed $(M,N)=(2,3)$, $K \in \{5,7\}$, and $\mathcal D(a,b) \in \{\mathcal U(-\frac{\pi}{2},\frac{\pi}{2}),\mathcal U(-\frac{\pi}{4},\frac{\pi}{6}), \mathcal TN(-\frac{\pi}{8},\frac{\pi}{8},0,1)\}$.    
We observe that the proposed MMSE combining approach attains superior MSE in estimating $\mathbf Z$ for any distribution and support set considered for the DoAs in $\Theta$. Moreover, we notice that the averaging combining approach consistently outperforms the selection combining approach in accordance with our theoretical finding that averaging combining attains superior performance compared to selection combining. Moreover, we observe that the performance advantage of the proposed MMSE combining approach over its standard counterparts is more profound when $\theta_k\sim \mathcal U(-\frac{\pi}{4},\frac{\pi}{6})$. 
Last, we notice that for $\theta_k\sim \mathcal U(-\frac{\pi}{4},\frac{\pi}{6})$, the performance gap between the proposed MMSE combining and the standard averaging approach is greater for $K=7$.

\begin{figure}[!t]
	\centering
	\includegraphics[width=\linewidth]{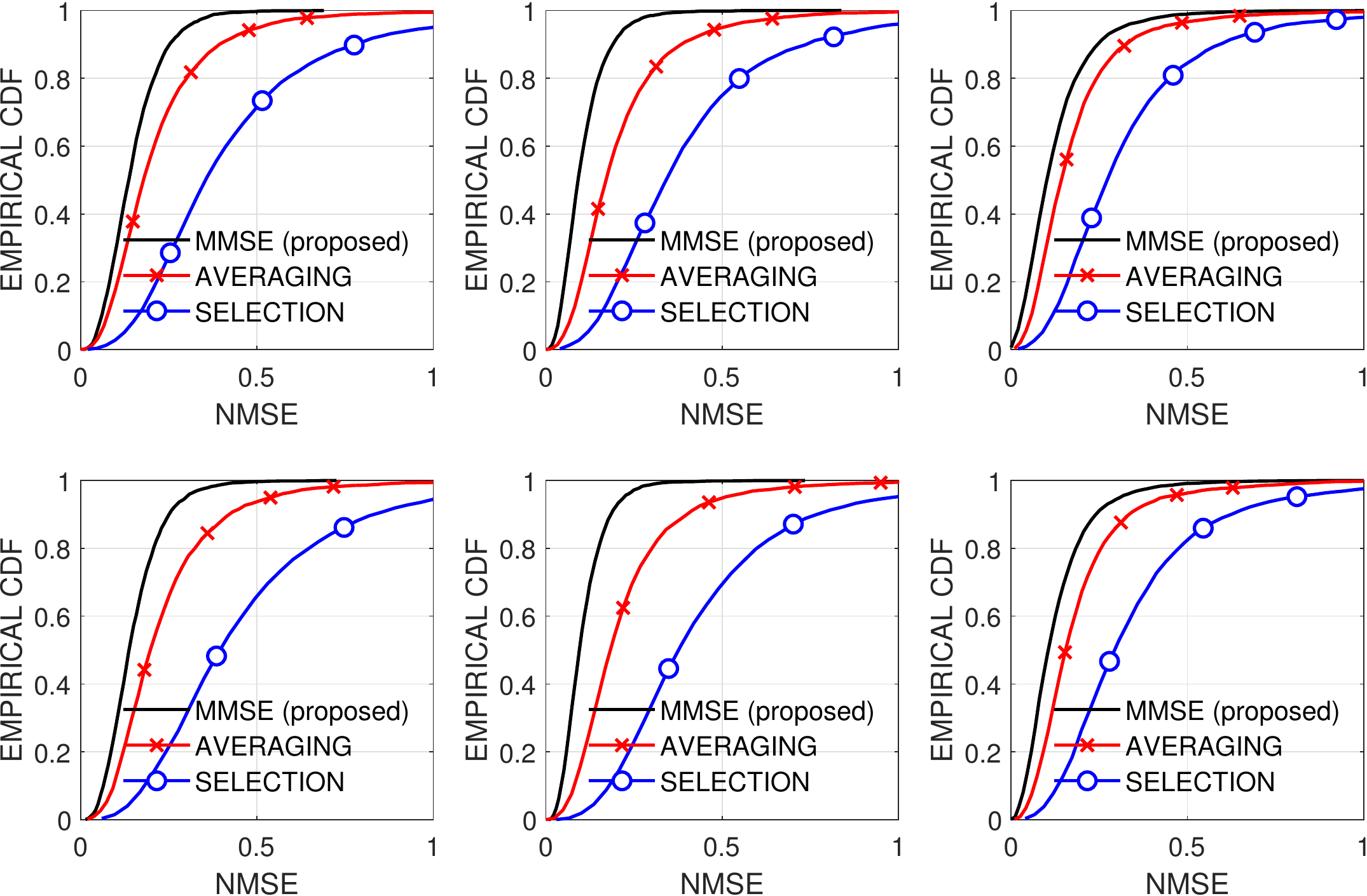}
	\caption{Empirical CDF of the NMSE in estimating $\mathbf Z_{}$ for $(M,N)=(2,5)$, SNR$=10$dB, $Q=10$,  $K=7$ (top), and $K=9$ (bottom). $\forall k, \theta_k\sim\mathcal U(-\frac{\pi}{2},\frac{\pi}{2})$ (left), $\mathcal U(-\frac{\pi}{4},\frac{\pi}{6})$ (center), and $\mathcal{TN}(-\frac{\pi}{8}, \frac{\pi}{8},0,1)$ (right).}
	\label{fig:cdf2}
\end{figure}

We repeat the last study for $(M,N)=(2,5)$, $K \in \{7,9\}$, and $\mathcal D(a,b) \in \{\mathcal U(-\frac{\pi}{2},\frac{\pi}{2}),\mathcal U(-\frac{\pi}{4},\frac{\pi}{6}), \mathcal TN(-\frac{\pi}{8},\frac{\pi}{8},0,1)\}$. We illustrate the new CDFs in Fig. \ref{fig:cdf2}.  
Similar observations as in Fig. \ref{fig:cdf1} are made. The proposed MMSE combining approach clearly outperforms its standard counterparts for any distribution assumption and support set for $\theta_k \forall k$.

\begin{figure}[!t]
	\centering
	\includegraphics[width=\linewidth]{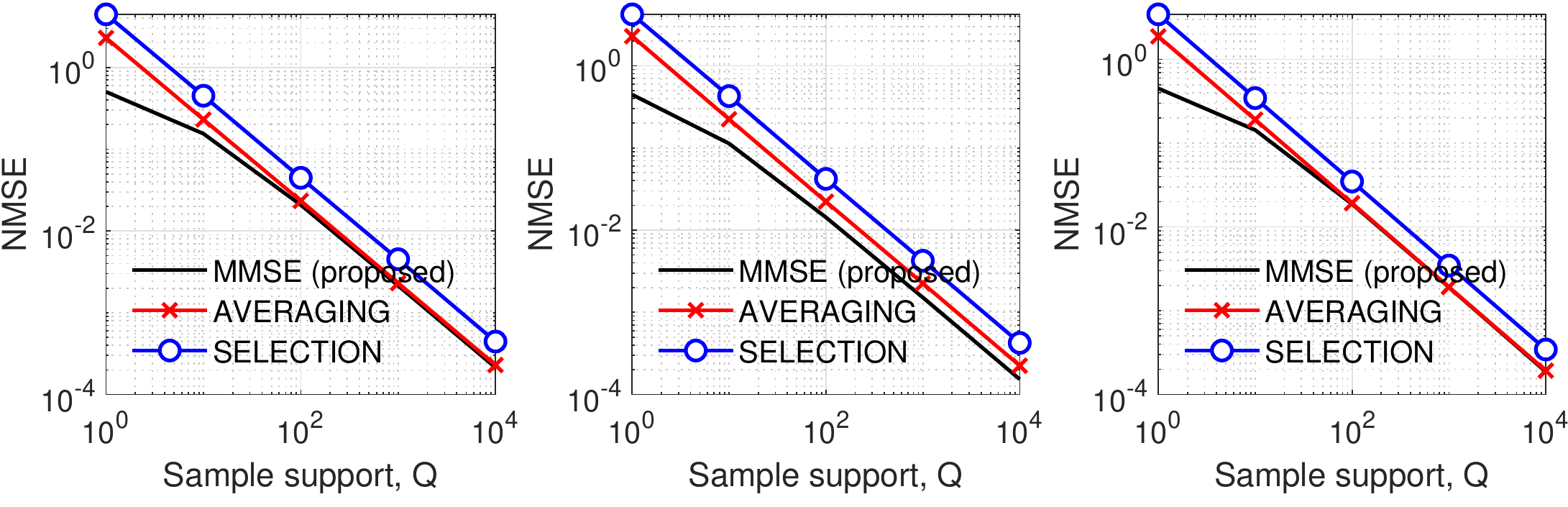}
	\caption{NMSE in estimating $\mathbf Z_{}$ versus sample support $Q$. $(M,N)=(2,5)$, SNR$=10$dB, and $K=7$. $\forall k, \theta_k\sim\mathcal U(-\frac{\pi}{2},\frac{\pi}{2})$ (left), $\mathcal U(-\frac{\pi}{4},\frac{\pi}{6})$ (center), $\mathcal{TN}(-\frac{\pi}{8},\frac{\pi}{8},0,1)$ (right).}
	\label{fig:mse}
\end{figure}

For $K=7$ and every other parameter same as above, we plot the NMSE (averaged over 4000 realizations) versus sample support $Q \in \{1, 10, 100, 1000, 10000\}$ in Fig. \ref{fig:mse}. 
Consistent with the observations above, we note that selection attains the highest NMSE while the proposed attains, expectedly, the lowest NMSE in estimating $\mathbf Z$ across the board. The performance gap between the proposed MMSE combining estimate and the estimates based on existing combining approaches decreases as the sample support $Q$ increases. Nonetheless, it remains superior, in many cases, even for high values of $Q$ (e.g., $Q=10^4$).

Moreover, in the first two subplots (uniform DoA distribution), we notice that the performance gap between the MMSE combining approach and the averaging approach is wider when the range of the support set $(a,b)$ is narrower.

\begin{figure}[!t]
	\centering
	\includegraphics[width=\linewidth]{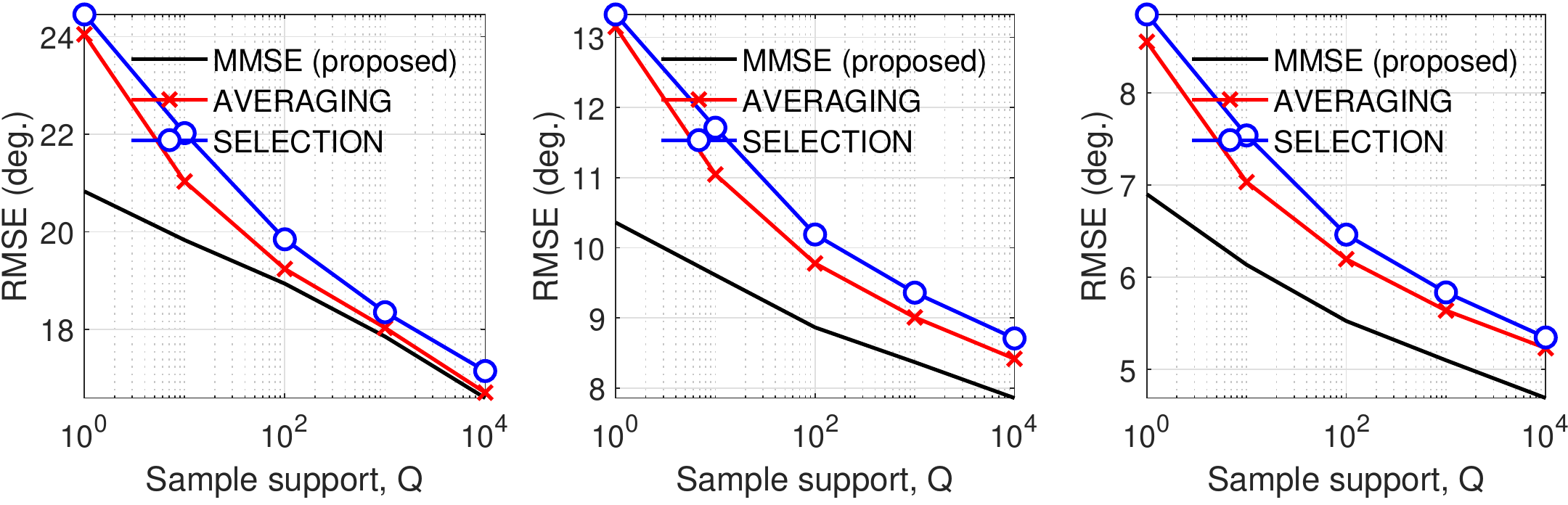}
	\caption{RMSE (degrees) in estimating the DoA set $\Theta$, versus sample support, $Q$.  $(M,N)=(2,5)$, SNR$=10$dB, and $K=7$. $\forall k, \theta_k\sim\mathcal U(-\frac{\pi}{2},\frac{\pi}{2})$ (left), $\mathcal U(-\frac{\pi}{4},\frac{\pi}{6})$ (center), $\mathcal{TN}(-\frac{\pi}{8},\frac{\pi}{8},0,1)$ (right).}
	\label{fig:rmse}
\end{figure}

We conclude the performance evaluations with measuring the Root-MSE (RMSE) in estimating the DoAs in $\Theta$ versus sample support $Q$. We first evaluate the MUSIC spectrum on each estimate $\hat{\mathbf Z} \in \{\hat{ \mathbf Z}_{sel}$, $\hat{ \mathbf Z}_{avg},\hat{ \mathbf Z}_{\text{MMSE}}\}$ over the probability distribution support set $(a, b)$ and obtain DoA estimates $\hat{\Theta}=\{\hat{\theta}_1,\ldots, \hat{\theta}_K\}$. 
Then, we compute the estimation error
\begin{align}
\text{MSE}(\hat{\Theta})=\frac{1}{K}\sum_{k=1}^K(\theta_k-\hat{\theta}_k)^2.
\end{align}
Finally, we compute the RMSE by taking the square root of the $\text{MSE}(\hat{\Theta})$ computed over 4000 statistically independent realizations of $\Theta$ and noise. The resulting RMSE curves are depicted in Fig. \ref{fig:rmse}. For every DoA distribution (even the most general case of uniform  distribution in $(-\frac{\pi}{2},\frac{\pi}{2})$) and every sample support (even as high as $10^4$), the proposed method attains the lowest RMSE. 

\section{Conclusions}
\label{SEC:CONCLUSIONS}
We proposed a novel coprime array receiver that attains minimum MSE in coarray autocorrelation estimation, for any probability distribution of the source DoAs. Moreover, we offered formal mathematical proofs for the closed-form MSE expressions of selection and averaging, which were first presented in \cite{DGC1}. Extensive numerical studies on various DoA distributions demonstrate that the proposed MMSE combining method consistently outperforms its existing counterparts in autocorrelation estimation performance with respect to the MSE metric. In turn, the proposed MMSE combiner enables lower RMSE in DoA estimation. 

\section*{Acknowledgments} 
This research is supported in part by the U.S. National Science Foundation under grant OAC-1808582 and U.S. Office of Naval Research under grants N00014-18-1-2460 and N00014-13-0061. This material is also based upon work supported by the Air Force Office of Scientific Research under award number FA9550-20-1-0039.

\section{Appendix}
\label{SEC: APPENDIX}
\subsection[Proof of Lemma 1.]{Proof of Lemma \ref{lem:lemma1}. }
\begin{proof}
We define $\mathbf b_q\eqdef \mathbf y_q^* \otimes \mathbf y_q$, for every $q=1,2,\ldots,Q$. 
Utilizing the the auxiliary variables\footnote{Recall that for any $x \in \mathbb N_+$, $\mathbf s_x=[1,2,\ldots, x]^T$.} $\dotv=\mathbf 1_L \otimes \mathbf s_L$, $\dotp=\mathbf p \otimes \mathbf 1_L$, $\ddotv=\mathbf s_L \otimes \mathbf 1_L$, and $\ddotp=\mathbf 1_L \otimes \mathbf p$, we find that 
$[\mathbf b_q]_{j_n}=[\mathbf y_q^*]_{[\dotv]_{j_n}}[\mathbf y_q]_{[\ddotv]_{j_n}}
$
for any $n \in \{1-L',2-L',\ldots L'-1\}$ and $j_n \in \mathcal{J}_n$,
with 
\begin{align}
&[\mathbf y_q]_{[\dotv]_{j_n}}=\sum_{k=1}^Kv(\theta_k)^{[\dotp]_{j_n}}\xi_{q,k}+[\mathbf n_q]_{[\dotv]_{j_n}},\\
&[\mathbf y_q]_{[\ddotv]_{j_n}}=\sum_{k=1}^Kv(\theta_k)^{[\ddotp]_{j_n}}\xi_{q,k}+[\mathbf n_q]_{[\ddotv]_{j_n}}.
\label{eq:gamma2}
\end{align} 
Next, we compute\footnote{Recall that for any $i \in \{1,2,\ldots,L^2\}$, $\omega_i= [\dotp]_i-[\ddotp]_i$.}
$\mathbb E\left\lbrace[\mathbf b_q]_{j_n}\right\rbrace=\sum_{k=1}^{K}v(\theta_k)^{\omega_{j_n}}d_k+\sigma^2[\mathbf i_L]_{j_n}=[\mathbf r]_{j_n}$. The latter implies that
$\mathbb E\left\lbrace\hat{\mathbf r}\right\rbrace=\mathbf r.$
We also define 
$
b_{q,p}^{(n)}(i,j)\eqdef[\mathbf b_{q}^*]_i[\mathbf b_{p}]_j,
$ 
for $(i,j) \in \mathcal{J}_n$ and $n \in \{1-L',2-L',\ldots, L'-1\}$ and compute
\begin{align}
\mathbb E \left\lbrace b_{q,p}^{(n)}(i,j)\right\rbrace=\Big|\mathbf g_n^H\mathbf d +\delta(n)\sigma^2\Big|^2+\delta(q-p)\Big|\mathbf z_{i,j}^H\mathbf d+\delta(i-j)\sigma^2\Big|^2,
\label{eq:gammanexp}
\end{align}
where $\mathbf g_n\eqdef\begin{bmatrix}
 v(\theta_1)^n & v(\theta_2)^n&\ldots&v(\theta_K)^n
\end{bmatrix}^\top$.
Next, we proceed as follows. 
\begin{align}
e&=\mathbb E\left\{ \Big|  [\mathbf r]_{j_n}  - [\hat{\mathbf r}]_{j_n}  \Big|^2 \right\}\\&=\Big|[\mathbf r]_{j_n}\Big|^2+\mathbb{E}\Big\lbrace[\hat{ \mathbf r}]_{j_n}[\hat{ \mathbf r}]_{j_n}^*\Big\rbrace -2\mathbb E \Big\lbrace  \Re\left\lbrace [ \mathbf r]_{j_n}[\hat{ \mathbf r}]_{j_n}^*\right\rbrace\Big\rbrace\\&=\mathbb{E}\Big\lbrace[\hat{ \mathbf r}]_{j_n}[\hat{ \mathbf r}]_{j_n}^*\Big\rbrace -\Big|[\mathbf r]_{j_n}\Big|^2\\&=\mathbb{E}\left\lbrace\frac{1}{Q^2}\sum_{q=1}^{Q}\sum_{p=1}^{Q}[ \mathbf b_q]_{j_n}[\mathbf b_p]_{j_n}^*\right\rbrace -\Big|[\mathbf r]_{j_n}\Big|^2
\\&=\frac{1}{Q^2}\sum_{q=1}^{Q}\sum_{p=1}^{Q}\mathbb E\left\lbrace b_{q,p}^{(n)}(j_n,j_n)\right\rbrace-\Big|[\mathbf r]_{j_n}\Big|^2
\\&\overset{\eqref{eq:gammanexp}}{=}\frac{(\mathbf 1_{K}^\top \mathbf d + \sigma^2)^2}{Q}.
\label{eq:e}
\end{align}
\end{proof}
\subsection[Proof of Lemma 2.]{Proof of Lemma \ref{lem:lemma2}. }
\begin{proof}
By Lemma \ref{lem:lemma1},  $e(\hat{\mathbf{r}}_{sel})=\mathbb E \left\lbrace\|\mathbf{r}_{co}-\hat{ \mathbf r}_{sel}\|_2^2\right\rbrace=\sum_{n=1-L'}^{L'-1}\mathbb E\left\{ \Big|  [\mathbf r]_{j_n}  - [\hat{\mathbf r}]_{j_n}  \Big|^2 \right\}=(2L'-1)e.$
\end{proof}

\subsection[Proof of Proposition 1.]{Proof of Proposition \ref{prop:prop1}.}
\begin{proof}
	Notice that 
	$
	\mathbf Z=\mathbf F(\mathbf I_{L'}\otimes \mathbf r_{co})=\big[
	\mathbf F_1  \mathbf r_{co} ~ \mathbf F_2  \mathbf r_{co} ~ \ldots ~\mathbf F_{L'}  \mathbf r_{co}
	\big].$
	By its definition, $\mathbf F_m$ is a selection matrix that selects the $\{L'-(m-1), L'-(m-2), \ldots, 2L'-m\}$-th entries of the length-$(2L'-1)$ vector it multiplies, for every $m \in \{1,2,\ldots, L'\}$. That is,  $\mathbf F_m \mathbf r_{co}=[\mathbf r_{co}]_{L'-(m-1):2L'-m}$.
	Similarly, $\hat{\mathbf Z}_{sel}=\big[
	\mathbf F_1  \hat{\mathbf r}_{sel} ~ \mathbf F_2  \hat{\mathbf r}_{sel} ~ \ldots ~\mathbf F_{L'}  \hat{\mathbf r}_{sel}\big]$ with $\mathbf F_m \hat{\mathbf r}_{sel}=[\hat{\mathbf r}_{sel}]_{L'-(m-1):2L'-m}$.
	In view of the above, 
	\begin{align}
	e(\hat{\mathbf Z}_{sel})&=\mathbb E\left\{ \left\| \mathbf Z  - \hat{\mathbf Z}_{sel}  \right\|_F^2 \right\}\\&=\mathbb E\left\{ \sum_{m=1}^{L'} \left\|\mathbf F_m \mathbf r_{co}-\mathbf F_m \hat{\mathbf r}_{sel}\right\|_2^2 \right\}\\&=\sum_{m=1}^{L'}\mathbb E\left\{  \left\|[\mathbf r_{co}]_{L'-(m-1):2L'-m}-[\hat{\mathbf r}_{sel}]_{L'-(m-1):2L'-m}\right\|_2^2 \right\}
	\\&=\sum_{m=1}^{L'} \sum_{n=1-m}^{L'-m}\mathbb E\left\{ \Big|[\mathbf r_{co}]_{L'+n}-[\hat{\mathbf r}_{sel}]_{L'+n}\Big|^2\right\}
	\\&=\sum_{m=1}^{L'} \sum_{n=1-m}^{L'-m}e
	\\&=L'^2e.
	\end{align} 
\end{proof} 
\subsection[Proof of Lemma 3.]{Proof of Lemma \ref{lem:lemma3}.}
\begin{proof}
\begin{align}
e_n&=\mathbb E \Big\{ \Big| [\mathbf r]_{j_n}-\frac{1}{|\mathcal{J}_n|}\sum_{j \in \mathcal J_n}[\hat{ \mathbf r}]_j\Big|^2 \Big\}\\&=\Big|[\mathbf r]_{j_n}\Big|^2+ \frac{1}{|\mathcal{J}_n|^2}\mathbb E \left\lbrace \Big| \sum_{j \in \mathcal{J}_n} [\hat{\mathbf r}]_j\Big|^2\right\rbrace-2\mathbb E \left\lbrace\Re \Bigg\{[\mathbf r]_{j_n}^*\left(\frac{1}{|\mathcal{J}_n|}\sum_{j \in \mathcal J_{n}}[\hat{\mathbf r}]_j\right)\Bigg\}\right\rbrace\\&=\Big|[\mathbf r]_{j_n}\Big|^2 +\frac{1}{|\mathcal{J}_n|^2}\sum_{j \in \mathcal{J}_n}\sum_{i \in \mathcal{J}_n}\frac{1}{Q^2  }\sum_{q=1}^{Q}\sum_{p=1}^{Q}\mathbb E \left\lbrace  b_{q,p}^{(n)}(i,j)\right\rbrace-2\Re \Bigg\{[\mathbf r]_{j_n}^*\left(\frac{1}{|\mathcal{J}_n|}\sum_{j \in \mathcal J_{n}}[\mathbf r]_j\right)\Bigg\}
\\&\overset{\eqref{eq:gammanexp}}{=}\frac{1}{Q}\left(\frac{2\sigma^2\mathbf 1_K^\top \mathbf d+\sigma^4}{|\mathcal{J}_n|}+\sum_{i \in \mathcal{J}_n}\sum_{j \in \mathcal J_n}\frac{|\mathbf z_{i,j}^H\mathbf d|^2}{|\mathcal{J}_n|^2}\right).
\label{eq:en}
\end{align} 
\end{proof}

\subsection[Proof of Lemma 4.]{Proof of Lemma \ref{lem:lemma4}. }
\begin{proof}
By Lemma \ref{lem:lemma3},  $e(\hat{\mathbf{r}}_{avg})=\mathbb E \left\lbrace\|\mathbf{r}_{co}-\hat{ \mathbf r}_{avg}\|_2^2\right\rbrace=\sum_{n=1-L'}^{L'-1}\mathbb E\left\{ \left|  [\mathbf r]_{j_n}  - \frac{1}{|\mathcal{J}_n|}\sum_{j \in \mathcal{J}_n}[\hat{\mathbf r}]_{j}  \right|^2 \right\}=\sum_{n=1-L'}^{L'-1}e_n$. 
\end{proof}

\subsection[Proof of Proposition 2.]{Proof of Proposition \ref{prop:prop2}.}
\begin{proof}
We know that
$
\mathbf Z=\big[
\mathbf F_1  \mathbf r_{co} ~ \mathbf F_2  \mathbf r_{co} ~ \ldots ~\mathbf F_{L'}  \mathbf r_{co}
\big].
$
Similarly,  $
\hat{\mathbf Z}_{avg}=\big[
\mathbf F_1  \hat{\mathbf r}_{avg} ~ \mathbf F_2  \hat{\mathbf r}_{avg} ~ \ldots ~\mathbf F_{L'}  \hat{\mathbf r}_{avg}
\big].
$
By the definition of $\mathbf F_m$, for every $m \in \{1,2,\ldots, L'\}$ it holds that  $\mathbf F_m \mathbf r_{co}=[\mathbf r_{co}]_{L'-(m-1):2L'-m}$ and $\mathbf F_m \hat{\mathbf r}_{avg}=[\hat{\mathbf r}_{avg}]_{L'-(m-1):2L'-m}$.
In view of the above, 
\begin{align}
e(\hat{\mathbf Z}_{avg})&=\mathbb E\left\{ \left\| \mathbf Z  - \hat{\mathbf Z}_{avg}  \right\|_F^2 \right\}\\&=\mathbb E\left\{ \sum_{m=1}^{L'} \left\|\mathbf F_m \mathbf r_{co}-\mathbf F_m \hat{\mathbf r}_{avg}\right\|_2^2 \right\}\\&=\sum_{m=1}^{L'}\mathbb E\left\{  \left\|[\mathbf r_{co}]_{L'-(m-1):2L'-m}-[\hat{\mathbf r}_{avg}]_{L'-(m-1):2L'-m}\right\|_2^2 \right\}
\\&=\sum_{m=1}^{L'} \sum_{n=1-m}^{L'-m}\mathbb E\left\{ \Big|[\mathbf r_{co}]_{L'+n}-[\hat{\mathbf r}_{avg}]_{L'+n}\Big|^2\right\}
\\&=\sum_{m=1}^{L'} \sum_{n=1-m}^{L'-m}e_n.
\end{align} 
\end{proof}
\subsection[Proof of Lemma 5.]{Proof of Lemma \ref{lem:w_statistics}.}
\label{SEC: PROOF_w_statistics}
We recall that $\mathbf w=\frac{1}{Q}\sum_{q=1}^{Q}\mathbf x_q^* \otimes \mathbf x_q$. Next, we notice that by utilizing the auxiliary variables\footnote{Recall that for any $x \in \mathbb N_+$, $\mathbf s_x=[1,2,\ldots, x]^T$.} $\dot{\mathbf u}=\mathbf 1_{K+L} \otimes \mathbf s_{K+L}$ and $\ddot{\mathbf u}=\mathbf s_{K+L}\otimes \mathbf 1_{K+L}$, we obtain
$
[\mathbf w]_i=\frac{1}{Q}\sum_{q=1}^{Q}[\mathbf x_q^*]_{[\ddot{\mathbf u}]_i}[\mathbf x_q]_{[\dot{\mathbf u}]_i}$.
Then, we define $\mathcal{I}\eqdef\big\{i \in \{1,2,\ldots,(K+L)^2\}: [\mathbf i]_i=1\big\}$, and observe that 
$
\mathbb E \left\lbrace[\mathbf x_q^*]_{[\ddot{\mathbf u}]_i}[\mathbf x_q]_{[\dot{\mathbf u}]_i}\right\rbrace=\delta([\dot{\mathbf u}]_i-[\ddot{\mathbf u}]_i)$ which is equal to $1$, if $i \in \mathcal{I}$ and $0$ if $i \notin \mathcal{I}$. 
The latter implies that
\begin{align}
\mathbb E\{\mathbf w\}=\mathbf i.
\end{align}
Next, for $(i,m) \in \{1,2,\ldots,(K+L)^2\}^2$, we define
$\eta_{i,m} \eqdef [\mathbf x_q^*]_{[\ddot{\mathbf u}]_i}[\mathbf x_q]_{[\dot{\mathbf u}]_i}[\mathbf x_p^*]_{[\dot{\mathbf u}]_m}[\mathbf x_p]_{[\ddot{\mathbf u}]_m}$. 
It holds 
$
 [\mathbf w]_i [\mathbf w^*]_m=\frac{1}{Q^2} \sum_{q=1}^{Q}\sum_{p=1}^{Q}\eta_{i,m}.
$
Exploiting the $2$-nd and $4$-th order moments of zero-mean independent normal variables, we find that
$
\mathbb{E}\left\lbrace\eta_{i,m}\right\rbrace$ is equal to
$1 +\delta(p-q)\delta(i-m)$ if $ (i,m) \in \mathcal{I},$ and $0$ otherwise.
The latter implies that
$
\mathbb E\{[\mathbf w]_i [\mathbf w^*]_m\}=\frac{1}{Q^2} \sum_{q=1}^{Q}\sum_{p=1}^{Q}\mathbb E \{\eta_{i,m}\}$ is equal to
$1+\frac{1}{Q}\delta(i,m)$, if $(i,m) \in \mathcal{I}$ and $0$ otherwise.
Altogether, we have
\begin{align}
\mathbb E_{\mathbf w} \{\mathbf w\mathbf w^H\}=\mathbf i \mathbf i^\top +\frac{1}{Q}\mathbf I_{K+L}.
\end{align}
\subsection[Proof of Lemma 6.]{Proof of Lemma \ref{lemmaHe}.}
\label{SEC: PROOF_HE}
\begin{proof}

For $(i,m) \in \{1,2,\ldots,L^2\}^2$,
$
[\mathbf H]_{i,m}=[\mathbf V \mathbf w]_i[(\mathbf V\mathbf i)^*]_m 
=\sum_{j=1,l=1}^{(K+L)^2}[\mathbf V]_{i,j}[\mathbf w]_j [\mathbf V^*]_{m,l}[\mathbf i]_l.
$
Accordingly, 
$
[\mathbf H_{\mathbb E}]_{i,m}=\sum_{j=1,l=1}^{(K+L)^2} \mathbb E_{\Theta}\mathbb E_{\mathbf w}\left\lbrace[\mathbf V]_{i,j}[\mathbf w]_j [\mathbf V]_{m,l}^*[\mathbf i]_l\right\rbrace.
$
Considering that the random variables $\mathbf w$ and $\mathbf V$ are statistically independent from each other and that $\mathbb E_{\mathbf w}\{\mathbf w\}=\mathbf i$ (see Lemma \ref{lem:w_statistics}), we obtain
$
[\mathbf H_{\mathbb E}]_{i,m}=\mathbb E_{\Theta}\left\lbrace[\mathbf V \mathbf i]_i[(\mathbf V\mathbf i)^*]_m\right\rbrace=\mathbb E_{\Theta}\left\lbrace [\mathbf r]_i [\mathbf r^*]_m\right\rbrace.
$
Then, we  substitute\footnote{Recall that for any $i \in \{1,2,\ldots,L^2\}$, $\omega_i= [\dotp]_i-[\ddotp]_i$.} $
[\mathbf r]_i=\sum_{k=1}^{K}v(\theta_k)^{\omega_i}d_k+\sigma^2 [\mathbf i_L]_i
$  
in $\mathbb E_{\Theta}\left\lbrace[\mathbf r]_i[\mathbf r^*]_m\right\rbrace$
and perform plain algebraic operations, obtaining 
\begin{align}
[\mathbf H_{\mathbb E}]_{i,m}&=\left\|\mathbf d\right\|_2^2\mathcal{I}\left(\omega_i-\omega_m\right)+\sigma^4\delta(\omega_i)\delta(\omega_m)
+\sigma^2\left(\mathbf 1_K^\top\mathbf d\right)\Big(\delta(\omega_i)\mathcal{I}(-\omega_m)+\mathcal{I}(\omega_i)\delta(\omega_m)\Big)
\\&+\mathcal{I}(\omega_i)\mathcal{I}(-\omega_m)\left((\mathbf 1_K^\top\mathbf d)^2-\left\|\mathbf d\right\|_2^2\right).
\end{align}
\end{proof}
\subsection[Proof of Lemma 7.]{Proof of Lemma \ref{lemmaGe}.}
\label{SEC: PROOF_GE}
\begin{proof}
For $(i,m) \in \{1,2,\ldots,L^2\}^2$ it holds $
[\mathbf G]_{i,m}=[\mathbf V \mathbf w]_i [(\mathbf V \mathbf w)^*]_m=\sum_{j=1,l=1}^{(K+L)^2} [\mathbf V]_{i,j} [\mathbf w]_j [\mathbf V^*]_{m,l} [\mathbf w^*]_{l}.
$
Accordingly,
$
[\mathbf G_{\mathbb E}]_{i,m}=\sum_{j=1,l=1}^{(K+L)^2} \underset{\Theta}{\mathbb E}\underset{\mathbf w}{\mathbb E}  
\left\lbrace   [\mathbf V]_{i,j} [\mathbf w]_j [\mathbf V^*]_{m,l} [\mathbf w^*]_{l} \right\rbrace.
$
Next, we recall that the random variables $\Theta$ and $\mathbf w$ are statistically independent from each other. Thus,  
$
[\mathbf G_{\mathbb E}]_{i,m}=\sum_{j=1,l=1}^{(K+L)^2} \underset{\Theta}{\mathbb E}
\left\lbrace   [\mathbf V]_{i,j}  \underset{\mathbf w}{\mathbb E} \left\lbrace [\mathbf w]_j [\mathbf w^*]_l\right\rbrace [\mathbf V^*]_{m,l}  \right\rbrace.
$
The latter is equivalent to 
$
\mathbf G_{\mathbb E}=\underset{\Theta}{\mathbb E}
\left\lbrace \mathbf V \underset{\mathbf w}{\mathbb E} \left\lbrace \mathbf w \mathbf w^H\right\rbrace \mathbf V^H \right\rbrace.
$
Then, by Lemma \ref{lem:w_statistics} we obtain
$
\mathbf G_{\mathbb E}= \underset{\Theta}{\mathbb E}
\left\lbrace \mathbf V \mathbf i \mathbf i^\top \mathbf V^H +\frac{1}{Q}\mathbf V \mathbf V^H\right\rbrace.
$
By Lemma \ref{lemmaHe}, we find that
$
\mathbf G_{\mathbb E}= \mathbf H_{\mathbb E} +\frac{1}{Q}\tilde{\mathbf V}_{\mathbb E} $.
\end{proof}

\bibliographystyle{unsrt}  
\bibliography{mmse_coprime}

\end{document}